\documentclass[universe,article,accept,pdftex,moreauthors]{Definitions/mdpi} 

\firstpage{1} 
\pubvolume{11}
\issuenum{8}
\articlenumber{280}
\pubyear{2025}
\copyrightyear{2025}
\externaleditor{Andrea Lapi} 
\datereceived{7 July 2025} 
\daterevised{11 August 2025} 
\dateaccepted{12 August 2025} 
\datepublished{21 August 2025} 
\hreflink{https://\linebreak doi.org/10.3390/universe11080280} 

\Title{Spectroscopic Observations and Emission-Line Diagnoses  for {H\,\pmb{\sc ii}}~Regions in the Late-Type Spiral Galaxy NGC\,2403}

\TitleCitation{Spectroscopic Observations and Emission-Line Diagnoses  for \HII~Regions in the Late-Type Spiral Galaxy NGC\,2403}

\Author{{Qi-Ming} 
 Wu $^{1,2}$\orcidA{}, Ye-Wei Mao $^{1,2,}$*\orcidB{}, Lin Lin $^{3}$\orcidC{}, Hu Zou $^{4}$\orcidD{} and Shu-Ting Wang $^{1, 2}$}

\AuthorNames{Qi-Ming Wu, Ye-Wei Mao, Lin Lin, Hu Zou and Shu-Ting Wang}

  \AuthorCitation{Wu, Q.-M.; Mao, Y.-W.; Lin, L.; Zou, H.; Wang, S.-T.}

\address{%
$^{1}$ \quad Center for Astrophysics, {Guangzhou} University, {Guangzhou} 510006, China; {qimingwu@e.gzhu.edu.cn~(Q.-M.W.); wangshuting@e.gzhu.edu.cn (S.-T.W.)} 
 \\
$^{2}$ \quad Department of Astronomy, School of Physics and Materials Science, {Guangzhou} University, {Guangzhou}~510006, China \\
$^{3}$ \quad Shanghai Astronomical Observatory, Chinese Academy of Sciences, {Shanghai} 200030, China; {linlin@shao.ac.cn} \\
$^{4}$ \quad National Astronomical Observatories, Chinese Academy of Sciences, {Beijing} 100101, China; {zouhu@nao.cas.cn}
}

\corres{Correspondence: ywmao@gzhu.edu.cn}

\usepackage{color}
\usepackage{aas_macros}
\usepackage{threeparttable}
\newcommand{\nodata}{--}
\newcommand{\OII}{\hbox{[O\,{\sc ii}]}}

\newcommand{\OIII}{\hbox{[O\,{\sc iii}]}}
\newcommand{\NII}{\hbox{[N\,{\sc ii}]}}
\newcommand{\SII}{\hbox{[S\,{\sc ii}]}}
\newcommand{\SIII}{\hbox{[S\,{\sc iii}]}}

\newcommand{\HII}{\hbox{H\,{\sc ii}}}

\newcommand{\Ha}{\hbox{H$\alpha$}}
\newcommand{\Hb}{\hbox{H$\beta$}}

\abstract{Being ionized nebulae where star formation events take place, \HII~regions are not only natural laboratories for studying physical processes of star formation and photoionization but also signatures reflecting evolution of their internal stellar populations and hosting galaxies.  In this paper, we present a comprehensive analysis of spectral emission-line data for \HII~regions in the nearby spiral galaxy NGC\,2403, aimed at gaining deep insight into underlying properties and evolution for the \HII~regions and the galaxy.  The spectroscopic data are obtained through observations with the 2.16\,m telescope at National Astronomical Observatories of China and a collection of published data in the literature.  Photoionization modeling is combined in the analysis for diagnosing the spectral features and interpreting the observational data with certain physical mechanisms.   Results of this work not only involve estimates of a set of parameters such as metallicity, the ionization parameter, etc., and evolution stages for the \HII~regions in NGC\,2403 but also reveal distinct characteristics of different spectral features and their sensitivities to specific parameters, which provides an instructive implication for proper usages of emission-line diagnostics for \HII~regions or galaxies nearby and far away.
}

\begin{document}

\section{Introduction}

Being gaseous nebulae irradiated by recently formed massive stars, \HII~regions are conspicuous sites where star formation activities accompanying photoionization processes occur, and~young stellar populations surrounded by ionized interstellar medium reside.  Spectra of \HII~regions contain a great number of emission lines, which serve as a useful tool for probing not only astrophysical mechanisms/properties in \HII~regions but also evolution history of host galaxies (\citet{Peimbert_2017, maiolino2019, Kewley2019}).

Metallicity is an important parameter derivable from emission-line features for \HII~regions.  It behaves like a signature of element enrichment with star formation, and~hence its spatial distributions are an imprint of chemical evolution for galaxies.  The~most accurate method for deriving metallicity from spectral emission lines is based on estimates of electron temperature ($T_\mathrm{e}$) and referred to as the $T_\mathrm{e}$ or direct method (\mbox{\citet{Peimbert_2017}}).  However, the $T_\mathrm{e}$ method requires measurements of auroral lines such as $\OIII \lambda 4363$, $\NII \lambda 5755$, or~$\SIII \lambda 6312$, which are intrinsically weak in spectra and even undetectable at high metallicity.  In~this case, strong collisionally excited emission lines such as $\OII\lambda3727$, $\OIII\lambda\lambda4959,5007$, $\NII\lambda\lambda6548,6583$, and~$\SII\lambda\lambda6717,6731$ are applied to estimating metallicity with empirical or theoretical calibrations, and~they have provided more widely used diagnostics in studies of external galaxies (\citet{Alloin1979A&A....78..200A, Pagel10.1093/mnras/189.1.95, PP2004, Pilyugin_Thuan2005ApJ...631..231P, Bresolin2007}).

In addition to metallicity, the~collisionally excited emission lines have been found to also relate with the ionization parameter and electron density at different levels of sensitivities.  Some spectral indices combining certain emission lines have been adopted to probe the physical parameters for \HII~regions or emission-line galaxies (e.g., $\OIII\lambda5007/\OII\lambda3727$ or $\SIII\lambda\lambda9069,9531/\SII\lambda\lambda6717,6731$ for probing the ionization parameter, \citet{KD2002, Morisset2016A&A...594A..37M}; $\OII\lambda3726/\OII\lambda3729$ or $\SII\lambda6717/\SII\lambda6731$ for probing electron density, \citet{Wang2004AA...427..873W}).\endnote{$\OII\lambda3727$ {is} 
 a sum of the double lines $\OII\lambda3726/\OII\lambda3729$, which are hardly resolvable for low-resolution spectroscopy and thus taken as one single line $\OII\lambda3727$.}

Since \HII~regions are excited by young massive stars, aging of the ionizing sources is considered to have an effect on the excitation and the subsequent spectral features (\citet{Levesque2010, Xiao2018MNRAS.477..904X}).  {Variations of commonly adopted diagnostic indices with stellar population age can be non-negligible, as~they may affect the inferred metallicity, ionization parameter, electron density, and~other physical properties. Nevertheless, the~impact of age on the emission lines of \HII regions has yet to be systematically addressed in observational studies of galaxies.}

{
\citet{Kong2014} conducted a spectroscopic survey of 20 nearby galaxies and obtained a substantial collection of \HII~region spectra. These data provide critical information on metallicity, ionization parameters, and~dust extinction, offering observational constraints for theoretical models of galaxy formation, chemical evolution, and~photoionization.  As~one of the target galaxies in this project,} NGC\,2403 is a late-type spiral galaxy (classified into the SAB(s)cd morphological type) located in the M81 galaxy group at the distance $\sim$$3.18$\,Mpc from the Milky Way (\citet{Tully2013AJ....146...86T}).~Basic information on astrometry for this galaxy includes the celestial coordinates R.A. = 07:36:51.3, Dec. = +65:36:09.7 (\citet{Skrutskie2006AJ....131.1163S_2mass}), the~apparent size $R_{25} =$ $\sim${$10^{\prime}.93$,} 
 the~inclination $\sim$$60^\circ$, and~the position angle {$\sim$$126^\circ$} (\citet{deVaucouleurs1991rc3..book.....D}).  There are a {number} of bright \HII~regions widespread in NGC\,2403, making it possible to conduct a spatially resolved spectroscopic analysis. {\citet{Mao2018} performed a spatially resolved study of 11 large, bright \HII~regions in NGC\,2403 using the \citet{Kong2014} dataset. They examined the performance of various metallicity calibration indicators within individual \HII~regions.}

\textls[15]{{In the present work,} we target \HII~regions in the late-type spiral galaxy NGC\,2403 and carry out an observational investigation in combination with modeling.  Spectroscopy is taken for the sampled \HII~regions and aimed at a comprehensive inspection of emission-line characteristics related to underlying physical properties.  In~particular, we attempt to derive the parameters from certain diagnostics on the one hand and scrutinize specific influences of certain parameters on different emission-line features on the other~hand.}

Subsequent content of this paper is structured as follows. Section~\ref{sec:Data} presents observations, data reduction, measurements, and~ancillary data from the literature.  Section~\ref{sec:Model} introduces the photoionization model employed in this study.  Results are described in \mbox{Section~\ref{sec:Result}}, followed by a discussion about implications and suggestions in Section~\ref{sec:Discussion}.  \mbox{Section~\ref{sec:Summary}} in the end summarizes conclusions drawn in this~work.

\section{Data}\label{sec:Data}

\subsection{Observations}

A total of 84 \HII~regions in NGC\,2403 were selected from the catalog of \mbox{\citet{Sivan1990}} and observed with the 2.16\,m telescope at the Xinglong Station of National Astronomical Observatories of China (NAOC) as part of a spectroscopic survey of \HII~regions in 20~nearby large galaxies (\citet{Kong2014}).  The~2.16\,m telescope equips the Optomechanics Research Inc. (OMR) long-slit spectrograph with a TEKTRONIX TEK1024, AR-coated back-illuminated CCD.  A~200\,{\AA}/mm dispersion grating, blazed at $\sim$$5500$\,{\AA}, was used in the spectrograph with a $4^{\prime}$-length slit, providing a dispersion sampling of 4.8\,{\AA} per pixel and a spectral resolution of 500--550 in $\lambda/\Delta\lambda$ at 5000\,{\AA} with a wavelength coverage of 3500--8000\,{\AA} (\citet{Fan2016}).

\begin{figure}[H]
	
        \includegraphics[width=\linewidth]{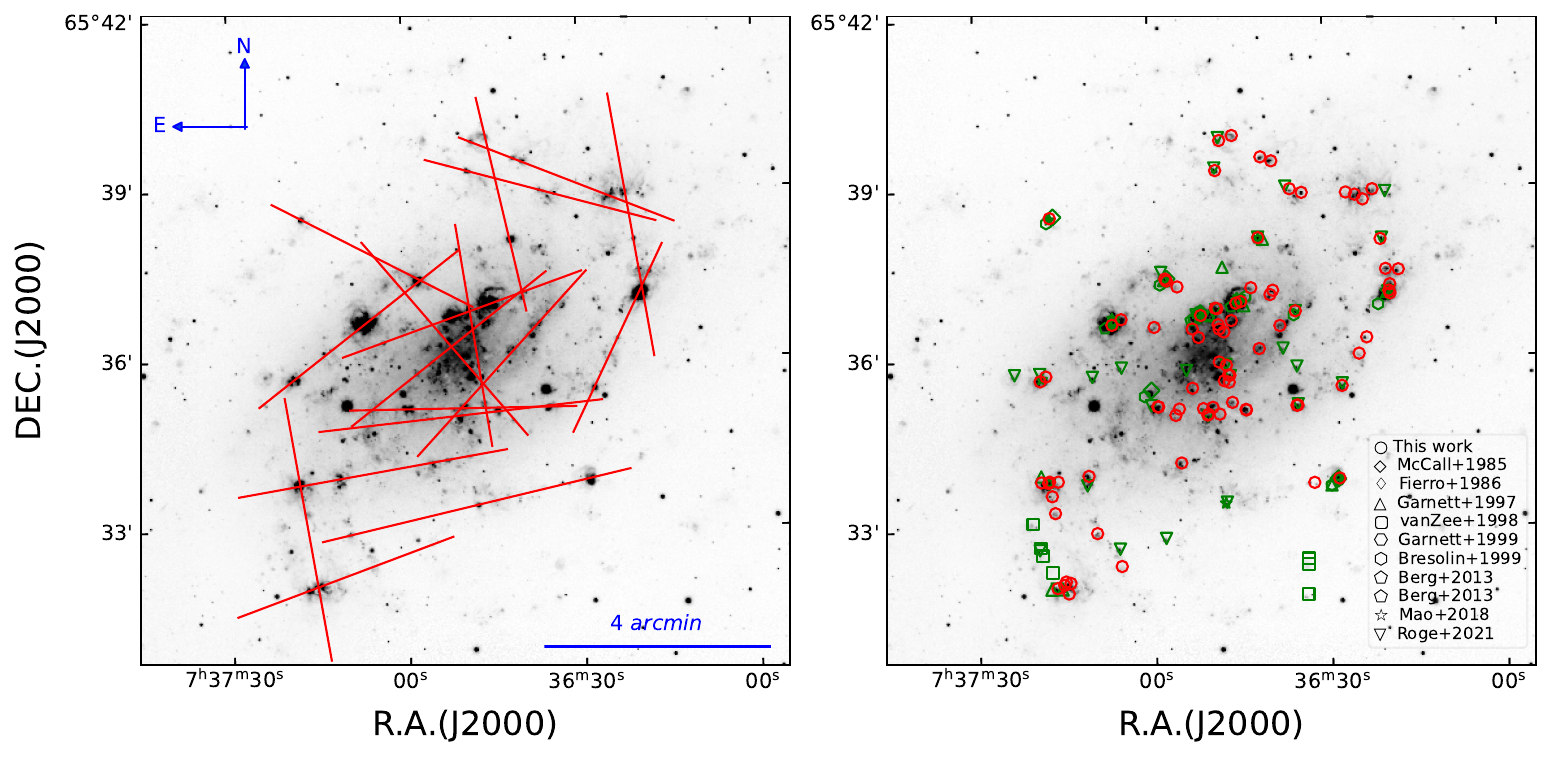}
	\caption{
    (\textbf{Left panel}): H$\alpha$ {narrow-band} 
 image for NGC\,2403 taken with the 2.1 m telescope at Kitt Peak National Observatory, with~the coverages of the spectrographic slit (red solid lines) superimposed. The~information of each of the slit coverages are listed in Table~\ref{table:slit}. 
    (\textbf{Right panel}): The same image as the left panel but with \HII~regions studied in this work marked by different symbols in different colors; the \HII~regions observed with the 2.16 m telescope at NAOC are marked with red open circles; those collected from other studies are colored in green and symbolized as listed in the lower-right corner of the panel. The~scale in the bottom right corner in the left panel indicates the 4~arcmin length which is the same in the right panel.  In~each of the panels, north is up, and east is to the~left.}
	\label{fig:Hii_slit}
\end{figure}

The long-slit spectroscopic observations were carried out over several nights from January 2007 to December 2008 (\citet{Kong2014, Mao2018}).  During~the observing runs, the~slit width was set to {$2^{\prime\prime}.5$} to match the average local seeing disk.  The~position angle of the long slit was oriented to cross as many \HII~regions as possible at one exposure frame.  The~exposure time for each slit position was 1800\,s $\times$ 2  or 1200\,s $\times$ 3 to achieve a suitable signal-to-noise (S/N) ratio.  Instrument bias and dome flats were recorded at the beginning and the end of each night.  A~He-Ar lamp was observed immediately after observing each object at the same position for wavelength calibration.  Spectrophotometric standard stars including Feige 34 and He 3 (\citet{Corbin1991irsc.book.....C}) were observed in each night for flux calibration.  Table~\ref{table:slit} lists statistics of the observations.  The~orientations of the spectrographic slit and the locations of the \HII~regions are shown in Figure~\ref{fig:Hii_slit}. 

\begin{table}[H]
	
	\caption{{Observations.} 
}\label{table:slit}

\begin{minipage}{\fulllength}
\begin{adjustwidth}{-\extralength}{0cm}
	\newcolumntype{C}{>{\centering\arraybackslash}X}
\begin{tabularx}{\textwidth}{CCCCCCC}
	
	\toprule
	
	\textbf{Slit ID.} & \textbf{Obs. Date} & \textbf{Slit Angle} & \textbf{R.A.} & \textbf{Decl.} & \textbf{Exp. Time} & \textbf{Airmass} \\
	\textbf{(1)} & \textbf{(2) }& \textbf{(3)} &\textbf{(4) }&\textbf{(5)} & \textbf{(6) }& \textbf{(7) }\\
        \midrule
	NGC\,2403-A & {\mbox{16 January 2007}} 
 & $-$33.7 & 07:37:00 & +65:36:08 & 1800 $\times$ 2 & 1.1128\\
	NGC\,2403-C & {\mbox{17 January 2007}} & +40.4 & 07:36:52 & +65:36:11 & 1800 $\times$ 2 & 1.1056\\
	NGC\,2403-E & {\mbox{18 January 2007}} & $-$52.0 & 07:37:07 & +65:36:39 & 1800 $\times$ 2 & 1.1059\\
	NGC\,2403-G & {\mbox{18 January 2007}} & $-$42.5 & 07:36:42 & +65:36:02 & 1800 $\times$ 2 & 1.1067\\
	NGC\,2403-H & {\mbox{01 January 2008}} & +9.7 & 07:36:21 & +65:38:01 & 1800 $\times$ 2 & 1.1064\\
	NGC\,2403-I & {\mbox{01 January 2008}} & +9.7 & 07:37:17 & +65:32:42 & 1800 $\times$ 2 & 1.1588\\
	NGC\,2403-O & {\mbox{02 January 2008}} & +68.4 & 07:36:32 & +65:39:09 & 1800 $\times$ 2 & 1.1218\\
	NGC\,2403-L & {\mbox{03 January 2008}} & $-$80.2 & 07:37:03 & +65:34:02 & 1800 $\times$ 2 & 1.1158\\
	NGC\,2403-K & {\mbox{04 January 2008}} & $-$83.9 & 07:36:50 & +65:35:01 & 1800 $\times$ 2 & 1.1118\\
	NGC\,2403-N & {\mbox{04 January 2008}} & $-$77.0 & 07:36:48 & +65:33:25 & 1800 $\times$ 2 & 1.1422\\
	NGC\,2403-Q & {\mbox{05 January 2008}} & +74.5 & 07:36:41 & +65:39:06 & 1800 $\times$ 2 & 1.1127\\
	NGC\,2403-S & {\mbox{06 January 2008}}& $-$26.1 & 07:36:23 & +65:36:19 & 1800 $\times$ 2 & 1.1055\\
	NGC\,2403-U & {\mbox{07 January 2008}} & $-$71.0 & 07:36:51 & +65:36:45 & 1800 $\times$ 2 & 1.1218\\
	NGC\,2403-01 & {\mbox{29 November 2008}} & $-$70.0 & 07:36:51 & +65:36:09 & 1200 $\times$ 3 & 1.1484\\
	NGC\,2403-04 & {\mbox{29 November 2008}} & +10.0 & 07:36:51 & +65:36:09 & 1200 $\times$ 3 & 1.1052\\
	NGC\,2403-05 & {\mbox{29 November 2008}} & +13.0 & 07:36:51 & +65:36:09 & 1200 $\times$ 3 & 1.1524\\
	NGC\,2403-02 & {\mbox{30 November 2008}} & +90.8 & 07:36:51 & +65:36:09 & 1200 $\times$ 3 & 1.1122\\
	NGC\,2403-03 & {\mbox{01 December 2008}} & +63.0 & 07:37:09 & +65:38:02 & 1200 $\times$ 3 & 1.1124\\
	\bottomrule
	\end{tabularx}
\end{adjustwidth}
\end{minipage}
\noindent{\footnotesize{{Notes.} 
 Columns: (1) the identifiers of the slit coverages; (2) the observation period; (3) the rotate angle; (4)--(5) R.A., Decl. for the central position of the slit; (6)--(7) the exposure time for each observation and air mass.}}

\end{table}%

\unskip

\subsection{Data~Reduction}

{Data reduction was conducted using PyRAF 2.2.0, a~Python-based interface to IRAF V2.17,\endnote{IRAF is distributed by the National Optical Astronomy Observatory, which is operated by the Association of Universities for Research in Astronomy, Inc., under~a cooperative agreement with the National Science Foundation.} employing the CCDRED, TWODSPEC, and~ONEDSPEC packages.}  The conventional routine including bias removal, flat-field correction, and~cosmic-ray rejection was implemented in the two-dimensional (2D) data in the form of a space--dispersion plane. 
{One-dimensional spectra were extracted using $10^{\prime\prime}$ apertures, corresponding to the $\sim$$154$\,pc local scale (by taking the distance $\sim$$3.18$\,Mpc addressed in \citet{Tully2013AJ....146...86T} into account). These apertures were placed along the spatial axis in the spatial–dispersion plane of the 2D spectra to extract the \HII~region profiles. When placing the apertures, we prioritized centering them on the peaks of the H$\alpha$ emission, provided that adjacent apertures would not overlap. In~cases where two nearby H$\alpha$ peaks were closely spaced, we slightly shifted the aperture centers to avoid overlap, while still ensuring coverage of the emission peaks.}

Spectral trajectories along the dispersion were traced by sampling and fitting continuum points with certain tracing function; the continuum points were obtained by summing enough dispersion lines; for some \HII~regions with very weak continua, spectra for the standard stars were used as reference for tracing the dispersion trajectories.  Background levels were estimated in blank areas along the spatial axis of the 2D data and subtracted from the spectra of the objects during the spectral extraction~process.

For each of the extracted spectra, the dispersion scale was calibrated to wavelength by referencing the emission-line spectra of the He-Ar lamp.  Flux calibration was performed by adopting the spectra of the spectrophotometric standard stars and the atmospheric extinction curve at the XingLong station. {During our observations, the~air mass ranged from 1.105 to 1.159 (mean 1.120). A~fixed {$2^{\prime\prime}.5$} slit width was adopted to match the average seeing of {$2^{\prime\prime}.0$} (\citet{Fan2016}). The~slit was not oriented at the parallactic angle in order to maximize target coverage. Following \citet{Filippenko1982PASP...94..715F}, we estimated differential atmospheric refraction (DAR) offsets at $\sec z \approx 1.15$ to be {$+0^{\prime\prime}.53$} at $[O II]\lambda 3727$ and {$-0^{\prime\prime}.27$} at $[S II]\lambda 6731$, relative to 5000\,\AA. At~$\sec z \approx 1.10$, the~offsets were {$+0^{\prime\prime}.43$} and {$-0^{\prime\prime}.22$}, respectively. Because~the slit width direction was not always perpendicular to the atmospheric dispersion direction, the~actual centroid shifts were smaller than the full DAR. At~3721\,\AA, all DAR-induced shifts were substantially smaller than both the local seeing and the slit width. Consequently, no DAR or slit-loss corrections were applied in this work.} All of the calibrated spectra were corrected for galactic foreground extinction, by~utilizing the \citet{Cardelli1989} (hereafter denoted as CCM) extinction curve and the total-to-selective extinction ratio $R_\mathrm{V} = 3.1$. The~color excess of the galactic extinction $E(B-V)_\mathrm{GAL} = 0.04$\,mag for NGC\,2403 was {obtained} from the \citet{Schlegel1998} galactic dust map and applied to the~correction.

\subsection{Emission-Line~Measurements}

The STARLIGHT software (\citet{Cid2005}) was employed to reproduce the underlying stellar continuum, for~the purpose of subtraction the continuum from each of the calibrated spectra to obtain sole emission-line spectra.  Before~the continuum fits, all of the observed spectra were corrected for redshift (shifted to the rest frame) and resampled to 1\,{\AA} per pixel using the SpectRes code { (\citet{Carnall2017arXiv170505165C})}, suiting the requirement of STARLIGHT.  The~model was composed of simple stellar populations (SSPs) selected from the \citet{Bruzual2003} spectral library of stellar population synthesis, encompassing 75 ages spanning from 1 Myr to 18 Gyr and 3 metallicities (Z = 0.008, 0.02, 0.05), on~the basis of the \citet{Chabrier2003} initial mass function (IMF) and the Padova 1994 evolutionary tracks (\citet{Le2003}).  Dust attenuation was imposed on the model spectra with the CCM extinction law  taken into account.  The~fits for the observed spectra were performed through a $\chi^2$-minimization approach using the Markov Chain Monte Carlo (MCMC) algorithm, and the best-fit model spectra were~obtained. 

Emission-line fluxes were measured in the continuum-subtracted spectra via Gaussian-profile fits.  Each of the emission lines $\OII\lambda3727$, H$\beta$, and~$\OIII\lambda5007$ was fitted with a single Gaussian profile; the three blended lines $\NII\lambda\lambda6548,6583$ + H$\alpha$ as well as the two blended lines $\SII\lambda\lambda6717,6731$ were deblended by fitting them with multiple Gaussian components.~The~flux ratio of $\OIII\lambda4959$ to $\OIII\lambda5007$ was fixed to 1:3, and so was $\NII\lambda6548$ to $\NII\lambda6583$.

Uncertainties of the emission-line fluxes were estimated using the expression in \citet{Gonzalez-Delgado1994ApJ...437..239G, Mao2018}{:} 
\begin{equation}
    \sigma_\mathrm{line} = \sigma_\mathrm{cont} N_\mathrm{pix}^{1/2}[1+EW/(N\Delta_{\lambda})]^{1/2} ,
\end{equation}
where $\sigma_\mathrm{cont}$ is the standard deviation of the continuum close to the emission line, $N_\mathrm{pix}$ is the number of pixels covering the measured emission line, $EW$ is the equivalent width of the emission line, and~$\Delta_{\lambda}$ is the dispersion in units of \AA/pixel.

The measured emission-line fluxes were corrected for dust attenuation using the Balmer decrement, following the prescription in \citet{Calzetti2001}:
\begin{equation}
   \frac{(L_{\Ha}/L_{\Hb})_{obs}}{(L_{\Ha}/L_{\Hb})_{ini}} = 10^{0.4E(B-V)[k(\Hb)-k(\Ha)]} ,
\end{equation}
where $(L_{\Ha}/L_{\Hb})_{obs}$ is the observed $\Ha$-to-$\Hb$ flux ratio, $(L_{\Ha}/L_{\Hb})_{ini} = 2.86$ is the assumed intrinsic $\Ha$-to-$\Hb$ flux ratio ratio in accordance with the Case B recombination at the electron temperature $T_e = 10^4~\rm K$ and the electron density $n_e = 100~\rm cm^{-3}$ (\citet{OsterbrockAstrophysics}), $k(\Hb)-k(\Ha) = 1.074$ and $R_\mathrm{V} = 3.1$ are quoted from the CCM extinction~law.  

The attenuation-corrected fluxes for the emission lines are listed in Table~\ref{table:line_ratio}.

\startlandscape

\begin{table}[H]
\caption{{Attenuation-corrected} 
 emission-line~fluxes.}
\footnotesize
\newcolumntype{C}{>{\centering\arraybackslash}X}
\begin{tabularx}{\textwidth}{cCCCCCCCCCc}

\toprule
\textbf{ID} & \textbf{R.A.} & \textbf{Dec.} & \textbf{R/R\boldmath{$_{25}$}} & \pmb{\Ha} & \pmb{$\OII\lambda3727$} &\pmb{$\OIII\lambda5007$} & \pmb{{$\NII\lambda6583$}} & \pmb{$\SII\lambda\lambda6717,31$} & \boldmath{$A_V$} & \pmb{\Hb}  \\
& \textbf{(J2000)}   & \textbf{(J2000)}   &   &   &   &   &   &    &   \textbf{(mag)}  & \boldmath{$10^{-15}\rm(erg\,s^{-1}\,cm^{-2})$} \\
\textbf{(1)} & \textbf{(2)} & \textbf{(3)} & \textbf{(4)} & \textbf{(5) }& \textbf{(6)} & \textbf{(7)} & \textbf{(8)} & \textbf{(9)} &\textbf{ (10)} & \textbf{(11) }\\
	\midrule
1 & 7:37:05.938 & +65:32:20.727 & 0.512 &  2.860 $\pm$ 0.804 & \nodata & 0.662 $\pm$ 0.213 & 0.288 $\pm$ 0.087 & 0.644 $\pm$ 0.175 & 0.559 $\pm$ 0.208  & 6.1  $\pm$  1.4\\
2 & 7:37:14.724 & +65:32:03.740 & 0.529 &  2.727 $\pm$ 0.018 & 2.699 $\pm$ 0.073 & 2.704 $\pm$ 0.019 & 0.265 $\pm$ 0.003 & 0.581 $\pm$ 0.010 & \nodata  & 35.4  $\pm$  0.2\\
3 & 7:37:15.797 & +65:32:01.689 & 0.532 &  2.843 $\pm$ 0.031 & 3.136 $\pm$ 0.143 & 1.493 $\pm$ 0.018 & 0.338 $\pm$ 0.007 & 0.723 $\pm$ 0.012 & \nodata  & 18.9  $\pm$  0.2\\
4 & 7:37:16.973 & +65:31:58.414 & 0.538 &  2.860 $\pm$ 0.123 & 2.413 $\pm$ 0.167 & 2.575 $\pm$ 0.125 & 0.292 $\pm$ 0.013 & 0.427 $\pm$ 0.019 & 0.385 $\pm$ 0.032  & 55.2  $\pm$  2.0\\
5 & 7:37:17.435 & +65:38:30.523 & 0.642 &  2.860 $\pm$ 0.113 & 1.840 $\pm$ 0.107 & 4.109 $\pm$ 0.184 & 0.098 $\pm$ 0.005 & 0.161 $\pm$ 0.011 & 0.239 $\pm$ 0.029  & 74.0  $\pm$  2.4\\
6 & 7:36:57.819 & +65:37:24.469 & 0.257 &  2.860 $\pm$ 0.070 & 3.503 $\pm$ 0.122 & 1.306 $\pm$ 0.036 & 0.467 $\pm$ 0.012 & 0.381 $\pm$ 0.013 & 0.750 $\pm$ 0.018  & 144.0  $\pm$  2.9\\
7 & 7:36:55.760 & +65:37:15.956 & 0.214 &  2.725 $\pm$ 0.083 & \nodata & 1.501 $\pm$ 0.046 & 0.710 $\pm$ 0.030 & 0.902 $\pm$ 0.038 & \nodata  & 5.8  $\pm$  0.2\\
8 & 7:36:35.620 & +65:35:08.406 & 0.332 &  2.860 $\pm$ 0.321 & 3.271 $\pm$ 0.518 & 1.506 $\pm$ 0.192 & 0.482 $\pm$ 0.055 & 0.689 $\pm$ 0.073 & 0.095 $\pm$ 0.083  & 43.2  $\pm$  4.0\\
9 & 7:36:44.249 & +65:35:04.334 & 0.240 &  2.860 $\pm$ 0.226 & 3.025 $\pm$ 0.530 & 0.866 $\pm$ 0.078 & 0.630 $\pm$ 0.051 & 0.730 $\pm$ 0.057 & 0.588 $\pm$ 0.059  & 48.8  $\pm$  3.2\\
10 & 7:36:49.959 & +65:35:07.739 & 0.174 &  2.860 $\pm$ 0.368 & 1.837 $\pm$ 0.468 & 0.850 $\pm$ 0.124 & 0.640 $\pm$ 0.083 & 0.821 $\pm$ 0.099 & 0.273 $\pm$ 0.095  & 27.2  $\pm$  2.9\\
11 & 7:36:51.597 & +65:35:06.539 & 0.163 &  2.860 $\pm$ 0.762 & \nodata & 0.523 $\pm$ 0.160 & 0.580 $\pm$ 0.156 & 0.810 $\pm$ 0.201 & 0.436 $\pm$ 0.198  & 18.3  $\pm$  4.0\\
12 & 7:36:59.240 & +65:35:09.214 & 0.130 &  2.860 $\pm$ 0.354 & 3.903 $\pm$ 0.682 & 0.609 $\pm$ 0.086 & 0.683 $\pm$ 0.085 & 0.812 $\pm$ 0.094 & 0.806 $\pm$ 0.092  & 51.8  $\pm$  5.3\\
13 & 7:36:59.389 & +65:35:09.230 & 0.130 &  2.860 $\pm$ 0.298 & 2.038 $\pm$ 0.386 & 0.645 $\pm$ 0.076 & 0.533 $\pm$ 0.057 & 0.520 $\pm$ 0.051 & 0.216 $\pm$ 0.077  & 35.9  $\pm$  3.1\\
14 & 7:36:49.283 & +65:36:52.725 & 0.100 &  2.860 $\pm$ 0.234 & 2.106 $\pm$ 0.260 & 0.771 $\pm$ 0.071 & 0.642 $\pm$ 0.053 & 0.722 $\pm$ 0.055 & 0.513 $\pm$ 0.061  & 36.6  $\pm$  2.5\\
15 & 7:36:48.835 & +65:36:35.144 & 0.055 &  2.860 $\pm$ 0.215 & 1.952 $\pm$ 0.212 & 0.712 $\pm$ 0.062 & 0.574 $\pm$ 0.044 & 0.621 $\pm$ 0.045 & 0.039 $\pm$ 0.056  & 15.8  $\pm$  1.0\\
16 & 7:36:48.686 & +65:36:30.517 & 0.045 &  2.540 $\pm$ 0.050 & 1.957 $\pm$ 0.180 & 0.459 $\pm$ 0.018 & 0.743 $\pm$ 0.019 & 0.959 $\pm$ 0.028 & \nodata  & 9.9  $\pm$  0.2\\
17 & 7:36:47.552 & +65:35:52.142 & 0.086 &  2.860 $\pm$ 0.326 & 1.943 $\pm$ 0.394 & 0.583 $\pm$ 0.077 & 0.892 $\pm$ 0.103 & 0.805 $\pm$ 0.086 & 0.795 $\pm$ 0.084  & 17.9  $\pm$  1.7\\
18 & 7:36:47.200 & +65:35:41.740 & 0.115 &  2.860 $\pm$ 0.193 & 3.699 $\pm$ 0.378 & 0.428 $\pm$ 0.034 & 0.818 $\pm$ 0.056 & 0.778 $\pm$ 0.050 & 1.137 $\pm$ 0.050  & 37.8  $\pm$  2.1\\
19 & 7:36:47.114 & +65:35:34.012 & 0.134 &  2.860 $\pm$ 0.413 & 2.272 $\pm$ 0.510 & 1.109 $\pm$ 0.182 & 0.815 $\pm$ 0.118 & 1.064 $\pm$ 0.144 & 0.169 $\pm$ 0.107  & 7.6  $\pm$  0.9\\
20 & 7:36:46.617 & +65:35:12.607 & 0.193 &  2.860 $\pm$ 0.319 & 2.091 $\pm$ 0.395 & 0.491 $\pm$ 0.063 & 0.527 $\pm$ 0.061 & 0.755 $\pm$ 0.078 & 0.321 $\pm$ 0.083  & 11.2  $\pm$  1.0\\
21 & 7:36:45.999 & +65:39:55.360 & 0.554 &  2.860 $\pm$ 0.722 & 2.232 $\pm$ 0.910 & 3.025 $\pm$ 0.862 & 0.239 $\pm$ 0.063 & 0.560 $\pm$ 0.131 & 0.585 $\pm$ 0.187  & 10.8  $\pm$  2.2\\
22 & 7:36:41.780 & +65:38:06.645 & 0.255 &  2.860 $\pm$ 0.087 & 1.787 $\pm$ 0.093 & 2.084 $\pm$ 0.072 & 0.264 $\pm$ 0.009 & 0.268 $\pm$ 0.011 & 0.066 $\pm$ 0.023  & 57.2  $\pm$  1.4\\
23 & 7:36:39.878 & +65:37:06.090 & 0.138 &  2.860 $\pm$ 0.168 & 1.954 $\pm$ 0.216 & 0.890 $\pm$ 0.060 & 0.567 $\pm$ 0.035 & 0.562 $\pm$ 0.032 & 0.119 $\pm$ 0.043  & 15.5  $\pm$  0.7\\
24 & 7:36:43.136 & +65:37:13.704 & 0.137 &  2.860 $\pm$ 0.208 & 3.837 $\pm$ 0.538 & 0.449 $\pm$ 0.038 & 0.656 $\pm$ 0.049 & 0.959 $\pm$ 0.065 & 0.434 $\pm$ 0.054  & 13.2  $\pm$  0.8\\
25 & 7:36:44.902 & +65:36:59.318 & 0.107 &  2.860 $\pm$ 0.078 & 3.091 $\pm$ 0.133 & 0.740 $\pm$ 0.023 & 0.513 $\pm$ 0.014 & 0.433 $\pm$ 0.013 & 0.197 $\pm$ 0.020  & 68.2  $\pm$  1.5\\
26 & 7:36:46.581 & +65:36:39.334 & 0.066 &  2.860 $\pm$ 0.137 & 3.918 $\pm$ 0.302 & 0.475 $\pm$ 0.026 & 0.723 $\pm$ 0.035 & 0.881 $\pm$ 0.039 & 0.628 $\pm$ 0.035  & 32.1  $\pm$  1.3\\
27 & 7:36:47.883 & +65:36:27.440 & 0.042 &  2.860 $\pm$ 0.319 & 4.171 $\pm$ 0.717 & 0.672 $\pm$ 0.088 & 0.761 $\pm$ 0.085 & 0.906 $\pm$ 0.096 & 0.761 $\pm$ 0.083  & 14.1  $\pm$  1.3\\
28 & 7:36:53.405 & +65:35:28.344 & 0.095 &  2.860 $\pm$ 0.256 & 3.081 $\pm$ 0.524 & 0.367 $\pm$ 0.041 & 0.826 $\pm$ 0.075 & 0.853 $\pm$ 0.072 & 0.946 $\pm$ 0.066  & 21.7  $\pm$  1.6\\
29 & 7:36:55.716 & +65:35:06.528 & 0.140 &  2.860 $\pm$ 0.364 & 4.103 $\pm$ 0.844 & 0.427 $\pm$ 0.067 & 0.753 $\pm$ 0.098 & 1.189 $\pm$ 0.146 & 0.115 $\pm$ 0.094  & 5.4  $\pm$  0.6\\
30 & 7:36:56.334 & +65:34:59.811 & 0.154 &  2.860 $\pm$ 0.472 & \nodata & 0.482 $\pm$ 0.091 & 0.927 $\pm$ 0.154 & 0.993 $\pm$ 0.153 & 0.809 $\pm$ 0.122  & 9.9  $\pm$  1.3\\
31 & 7:36:53.279 & +65:36:31.060 & 0.075 &  2.860 $\pm$ 0.347 & 4.779 $\pm$ 0.908 & 0.394 $\pm$ 0.057 & 0.701 $\pm$ 0.086 & 1.044 $\pm$ 0.117 & 1.129 $\pm$ 0.090  & 27.6  $\pm$  2.8\\
32 & 7:36:52.236 & +65:36:21.969 & 0.041 &  2.860 $\pm$ 0.195 & 2.085 $\pm$ 0.225 & 0.310 $\pm$ 0.026 & 0.793 $\pm$ 0.055 & 0.757 $\pm$ 0.049 & 0.173 $\pm$ 0.051  & 24.5  $\pm$  1.4\\
33 & 7:36:48.761 & +65:35:55.652 & 0.063 &  2.860 $\pm$ 0.325 & 2.823 $\pm$ 0.517 & 0.412 $\pm$ 0.056 & 1.022 $\pm$ 0.117 & 1.121 $\pm$ 0.117 & 0.582 $\pm$ 0.084  & 18.8  $\pm$  1.8\\
34 & 7:36:46.985 & +65:35:41.295 & 0.118 &  2.738 $\pm$ 0.049 & 2.062 $\pm$ 0.169 & 0.281 $\pm$ 0.011 & 0.935 $\pm$ 0.021 & 0.831 $\pm$ 0.022 & \nodata  & 17.9  $\pm$  0.3\\
35 & 7:36:57.586 & +65:37:22.390 & 0.250 &  2.785 $\pm$ 0.075 & 4.692 $\pm$ 0.203 & 0.610 $\pm$ 0.019 & 0.597 $\pm$ 0.018 & 0.824 $\pm$ 0.027 & \nodata  & 9.7  $\pm$  0.3\\
36 & 7:37:05.433 & +65:36:42.233 & 0.250 &  2.860 $\pm$ 0.070 & 2.593 $\pm$ 0.124 & 1.580 $\pm$ 0.044 & 0.489 $\pm$ 0.012 & 0.692 $\pm$ 0.017 & 0.246 $\pm$ 0.018  & 51.9  $\pm$  1.0\\
37 & 7:37:07.011 & +65:36:36.410 & 0.259 &  2.704 $\pm$ 0.008 & 2.689 $\pm$ 0.035 & 1.713 $\pm$ 0.008 & 0.406 $\pm$ 0.002 & 0.461 $\pm$ 0.009 & \nodata  & 205.0  $\pm$  0.6\\
38 & 7:37:18.452 & +65:35:42.768 & 0.330 &  2.860 $\pm$ 0.149 & 2.971 $\pm$ 0.289 & 1.920 $\pm$ 0.113 & 0.514 $\pm$ 0.027 & 0.593 $\pm$ 0.030 & 0.018 $\pm$ 0.039  & 13.5  $\pm$  0.6\\

    \bottomrule
\end{tabularx}
\end{table}%
\begin{table}[H]\ContinuedFloat
\caption{\em Cont.}
\footnotesize
\newcolumntype{C}{>{\centering\arraybackslash}X}
\begin{tabularx}{\textwidth}{cCCCCCCCCCc}

\toprule
\textbf{ID} & \textbf{R.A.} & \textbf{Dec.} & \textbf{R/R\boldmath{$_{25}$}} & \pmb{\Ha} & \pmb{$\OII\lambda3727$} &\pmb{$\OIII\lambda5007$} & \pmb{{$\NII\lambda6583$}} & \pmb{$\SII\lambda\lambda6717,31$} & \boldmath{$A_V$} & \pmb{\Hb}  \\
& \textbf{(J2000)}   & \textbf{(J2000)}   &   &   &   &   &   &    &   \textbf{(mag)}  & \boldmath{$10^{-15}\rm(erg\,s^{-1}\,cm^{-2})$} \\
\textbf{(1)} & \textbf{(2)} & \textbf{(3)} & \textbf{(4)} & \textbf{(5) }& \textbf{(6)} & \textbf{(7)} & \textbf{(8)} & \textbf{(9)} &\textbf{ (10)} & \textbf{(11) }\\
	\midrule

39 & 7:37:19.431 & +65:35:37.799 & 0.338 &  2.860 $\pm$ 0.154 & 3.354 $\pm$ 0.283 & 1.625 $\pm$ 0.099 & 0.515 $\pm$ 0.029 & 0.647 $\pm$ 0.034 & 0.713 $\pm$ 0.040  & 26.7  $\pm$  1.2\\
40 & 7:36:50.723 & +65:35:00.452 & 0.186 &  2.824 $\pm$ 0.024 & 1.532 $\pm$ 0.101 & 1.014 $\pm$ 0.011 & 0.461 $\pm$ 0.008 & 0.443 $\pm$ 0.013 & \nodata  & 17.1  $\pm$  0.1\\
41 & 7:36:35.603 & +65:36:48.393 & 0.174 &  2.860 $\pm$ 0.399 & \nodata & 1.044 $\pm$ 0.166 & 0.559 $\pm$ 0.079 & 0.659 $\pm$ 0.088 & 0.231 $\pm$ 0.103  & 13.8  $\pm$  1.6\\
42 & 7:36:38.275 & +65:36:33.111 & 0.149 &  2.777 $\pm$ 0.080 & 2.153 $\pm$ 0.207 & 1.151 $\pm$ 0.036 & 0.538 $\pm$ 0.027 & 0.768 $\pm$ 0.034 & \nodata  & 14.8  $\pm$  0.4\\
43 & 7:36:41.882 & +65:36:09.148 & 0.128 &  2.840 $\pm$ 0.120 & 1.611 $\pm$ 0.282 & 0.832 $\pm$ 0.040 & 0.646 $\pm$ 0.032 & 0.775 $\pm$ 0.038 & \nodata  & 11.6  $\pm$  0.5\\
44 & 7:36:47.955 & +65:35:35.674 & 0.121 &  2.860 $\pm$ 0.377 & 1.600 $\pm$ 0.360 & 0.523 $\pm$ 0.080 & 0.714 $\pm$ 0.095 & 0.516 $\pm$ 0.065 & 0.247 $\pm$ 0.098  & 22.0  $\pm$  2.4\\
45 & 7:36:22.076 & +65:38:56.202 & 0.382 &  2.860 $\pm$ 0.218 & 3.773 $\pm$ 0.474 & 1.184 $\pm$ 0.102 & 0.427 $\pm$ 0.034 & 0.679 $\pm$ 0.050 & 0.227 $\pm$ 0.056  & 23.1  $\pm$  1.4\\
46 & 7:36:20.847 & +65:38:03.288 & 0.339 &  2.860 $\pm$ 0.415 & 3.784 $\pm$ 0.867 & 1.852 $\pm$ 0.304 & 0.316 $\pm$ 0.047 & 0.484 $\pm$ 0.066 & 0.470 $\pm$ 0.108  & 24.4  $\pm$  2.9\\
47 & 7:36:20.019 & +65:37:31.521 & 0.345 &  2.827 $\pm$ 0.077 & 3.781 $\pm$ 0.242 & 0.754 $\pm$ 0.025 & 0.495 $\pm$ 0.021 & 0.842 $\pm$ 0.033 & \nodata  & 13.0  $\pm$  0.3\\
48 & 7:36:19.348 & +65:37:15.125 & 0.359 &  2.860 $\pm$ 0.113 & 3.270 $\pm$ 0.189 & 0.957 $\pm$ 0.043 & 0.448 $\pm$ 0.018 & 0.613 $\pm$ 0.028 & 0.134 $\pm$ 0.029  & 66.3  $\pm$  2.2\\
49 & 7:36:19.328 & +65:37:05.527 & 0.366 &  2.826 $\pm$ 0.009 & 2.308 $\pm$ 0.030 & 1.850 $\pm$ 0.009 & 0.307 $\pm$ 0.001 & 0.292 $\pm$ 0.009 & \nodata  & 199.0  $\pm$  0.6\\
50 & 7:37:18.249 & +65:33:49.977 & 0.334 &  2.860 $\pm$ 0.077 & 2.395 $\pm$ 0.104 & 2.013 $\pm$ 0.061 & 0.248 $\pm$ 0.007 & 0.355 $\pm$ 0.012 & 0.535 $\pm$ 0.020  & 124.0  $\pm$  2.8\\
51 & 7:37:17.668 & +65:33:35.847 & 0.350 &  2.860 $\pm$ 0.220 & 2.288 $\pm$ 0.307 & 1.006 $\pm$ 0.088 & 0.342 $\pm$ 0.030 & 0.655 $\pm$ 0.052 & 0.043 $\pm$ 0.057  & 12.2  $\pm$  0.8\\
52 & 7:37:17.171 & +65:33:17.894 & 0.377 &  2.521 $\pm$ 0.101 & 2.599 $\pm$ 0.231 & 4.768 $\pm$ 0.189 & 0.294 $\pm$ 0.020 & 0.531 $\pm$ 0.037 & \nodata  & 7.8  $\pm$  0.3\\
53 & 7:37:15.514 & +65:32:05.246 & 0.524 &  2.777 $\pm$ 0.038 & 3.783 $\pm$ 0.140 & 1.324 $\pm$ 0.019 & 0.323 $\pm$ 0.008 & 0.788 $\pm$ 0.020 & \nodata  & 20.0  $\pm$  0.3\\
54 & 7:37:15.053 & +65:31:52.534 & 0.554 &  2.762 $\pm$ 0.071 & 3.885 $\pm$ 0.243 & 1.680 $\pm$ 0.044 & 0.282 $\pm$ 0.015 & 0.521 $\pm$ 0.026 & \nodata  & 11.7  $\pm$  0.3\\
55 & 7:36:35.434 & +65:35:08.504 & 0.334 &  2.860 $\pm$ 0.110 & 3.844 $\pm$ 0.230 & 1.285 $\pm$ 0.056 & 0.449 $\pm$ 0.018 & 0.633 $\pm$ 0.025 & 0.386 $\pm$ 0.029  & 36.7  $\pm$  1.2\\
56 & 7:36:44.324 & +65:35:04.887 & 0.237 &  2.860 $\pm$ 0.181 & 3.610 $\pm$ 0.335 & 1.086 $\pm$ 0.078 & 0.586 $\pm$ 0.038 & 0.793 $\pm$ 0.049 & 0.322 $\pm$ 0.047  & 18.1  $\pm$  0.9\\
57 & 7:36:48.799 & +65:35:00.381 & 0.203 &  2.860 $\pm$ 0.443 & 4.695 $\pm$ 1.121 & 1.087 $\pm$ 0.192 & 0.641 $\pm$ 0.101 & 1.053 $\pm$ 0.152 & 0.908 $\pm$ 0.115  & 9.0  $\pm$  1.2\\
58 & 7:36:50.864 & +65:34:59.651 & 0.187 &  2.860 $\pm$ 0.243 & 1.904 $\pm$ 0.288 & 1.114 $\pm$ 0.108 & 0.557 $\pm$ 0.049 & 0.650 $\pm$ 0.052 & 0.460 $\pm$ 0.063  & 11.9  $\pm$  0.8\\
59 & 7:36:55.474 & +65:34:09.316 & 0.287 &  2.860 $\pm$ 0.315 & 3.012 $\pm$ 0.498 & 0.967 $\pm$ 0.121 & 0.535 $\pm$ 0.061 & 0.907 $\pm$ 0.097 & 0.079 $\pm$ 0.082  & 8.8  $\pm$  0.8\\
60 & 7:37:11.345 & +65:33:56.794 & 0.292 &  2.860 $\pm$ 0.280 & 4.006 $\pm$ 0.578 & 1.374 $\pm$ 0.153 & 0.456 $\pm$ 0.046 & 0.816 $\pm$ 0.076 & 0.625 $\pm$ 0.073  & 17.1  $\pm$  1.4\\
61 & 7:37:16.617 & +65:33:51.188 & 0.323 &  2.860 $\pm$ 0.385 & 4.502 $\pm$ 1.027 & 0.783 $\pm$ 0.120 & 0.410 $\pm$ 0.057 & 0.877 $\pm$ 0.109 & 0.895 $\pm$ 0.100  & 13.2  $\pm$  1.5\\
62 & 7:37:18.101 & +65:33:51.336 & 0.332 &  2.860 $\pm$ 0.088 & 2.665 $\pm$ 0.118 & 2.013 $\pm$ 0.070 & 0.253 $\pm$ 0.008 & 0.315 $\pm$ 0.013 & 0.371 $\pm$ 0.023  & 51.9  $\pm$  1.3\\
63 & 7:37:19.463 & +65:33:50.704 & 0.341 &  2.860 $\pm$ 0.177 & 2.634 $\pm$ 0.254 & 1.357 $\pm$ 0.096 & 0.369 $\pm$ 0.024 & 0.537 $\pm$ 0.032 & 0.314 $\pm$ 0.046  & 23.8  $\pm$  1.2\\
64 & 7:36:28.529 & +65:33:50.382 & 0.598 &  2.860 $\pm$ 0.066 & 2.633 $\pm$ 0.092 & 3.448 $\pm$ 0.091 & 0.173 $\pm$ 0.004 & 0.310 $\pm$ 0.010 & 0.090 $\pm$ 0.017  & 89.2  $\pm$  1.7\\
65 & 7:37:10.039 & +65:32:55.986 & 0.416 &  2.860 $\pm$ 0.245 & 3.895 $\pm$ 0.510 & 1.903 $\pm$ 0.184 & 0.410 $\pm$ 0.036 & 0.631 $\pm$ 0.052 & 0.595 $\pm$ 0.064  & 24.8  $\pm$  1.8\\
66 & 7:36:48.187 & +65:39:50.257 & 0.555 &  2.860 $\pm$ 0.198 & 2.440 $\pm$ 0.259 & 3.278 $\pm$ 0.255 & 0.147 $\pm$ 0.013 & 0.335 $\pm$ 0.024 & 0.188 $\pm$ 0.051  & 24.9  $\pm$  1.4\\
67 & 7:36:41.163 & +65:39:32.192 & 0.466 &  2.860 $\pm$ 0.291 & 3.236 $\pm$ 0.463 & 2.884 $\pm$ 0.331 & 0.223 $\pm$ 0.024 & 0.382 $\pm$ 0.037 & 0.104 $\pm$ 0.075  & 17.8  $\pm$  1.5\\
68 & 7:36:39.314 & +65:39:27.757 & 0.446 &  2.860 $\pm$ 0.208 & 3.157 $\pm$ 0.360 & 2.713 $\pm$ 0.222 & 0.237 $\pm$ 0.019 & 0.428 $\pm$ 0.031 & 0.051 $\pm$ 0.054  & 19.6  $\pm$  1.2\\
69 & 7:36:26.635 & +65:38:53.255 & 0.359 &  2.860 $\pm$ 0.491 & \nodata & 1.060 $\pm$ 0.206 & 0.514 $\pm$ 0.092 & 0.895 $\pm$ 0.164 & 0.182 $\pm$ 0.127  & 4.9  $\pm$  0.7\\
70 & 7:36:25.091 & +65:38:50.765 & 0.361 &  2.860 $\pm$ 0.184 & 2.989 $\pm$ 0.374 & 1.429 $\pm$ 0.104 & 0.414 $\pm$ 0.028 & 0.696 $\pm$ 0.044 & 0.490 $\pm$ 0.048  & 23.0  $\pm$  1.2\\
71 & 7:36:23.743 & +65:38:45.612 & 0.359 &  2.860 $\pm$ 0.289 & 2.961 $\pm$ 0.465 & 1.322 $\pm$ 0.151 & 0.537 $\pm$ 0.059 & 0.777 $\pm$ 0.075 & 0.034 $\pm$ 0.075  & 7.3  $\pm$  0.6\\
72 & 7:36:48.963 & +65:39:18.487 & 0.478 &  2.860 $\pm$ 0.182 & 2.953 $\pm$ 0.279 & 2.281 $\pm$ 0.164 & 0.255 $\pm$ 0.017 & 0.360 $\pm$ 0.025 & 0.456 $\pm$ 0.047  & 49.5  $\pm$  2.6\\
73 & 7:36:36.239 & +65:38:57.847 & 0.363 &  2.688 $\pm$ 0.085 & 3.557 $\pm$ 0.286 & 1.022 $\pm$ 0.034 & 0.623 $\pm$ 0.026 & 0.809 $\pm$ 0.042 & \nodata  & 9.3  $\pm$  0.3\\
74 & 7:36:34.204 & +65:38:53.771 & 0.351 &  2.706 $\pm$ 0.080 & \nodata & 1.151 $\pm$ 0.038 & 0.478 $\pm$ 0.019 & 1.020 $\pm$ 0.040 & \nodata  & 8.4  $\pm$  0.2\\
75 & 7:36:17.843 & +65:37:30.808 & 0.370 &  2.860 $\pm$ 0.198 & 3.711 $\pm$ 0.435 & 0.963 $\pm$ 0.077 & 0.410 $\pm$ 0.029 & 0.709 $\pm$ 0.048 & 0.167 $\pm$ 0.051  & 42.7  $\pm$  2.4\\
76 & 7:36:19.527 & +65:37:08.371 & 0.361 &  2.687 $\pm$ 0.015 & 2.442 $\pm$ 0.040 & 1.722 $\pm$ 0.010 & 0.332 $\pm$ 0.003 & 0.311 $\pm$ 0.009 & \nodata  & 178.0  $\pm$  0.9\\
77 & 7:36:23.465 & +65:36:19.340 & 0.361 &  2.860 $\pm$ 0.773 & 4.711 $\pm$ 2.049 & 1.598 $\pm$ 0.488 & 0.445 $\pm$ 0.123 & 0.704 $\pm$ 0.178 & 0.820 $\pm$ 0.200  & 17.1  $\pm$  3.8\\

    \bottomrule
\end{tabularx}
\end{table}%
\begin{table}[H]\ContinuedFloat
\caption{\em Cont.}
\footnotesize
\newcolumntype{C}{>{\centering\arraybackslash}X}
\begin{tabularx}{\textwidth}{cCCCCCCCCCc}

\toprule
\textbf{ID} & \textbf{R.A.} & \textbf{Dec.} & \textbf{R/R\boldmath{$_{25}$}} & \pmb{\Ha} & \pmb{$\OII\lambda3727$} &\pmb{$\OIII\lambda5007$} & \pmb{{$\NII\lambda6583$}} & \pmb{$\SII\lambda\lambda6717,31$} & \boldmath{$A_V$} & \pmb{\Hb}  \\
& \textbf{(J2000)}   & \textbf{(J2000)}   &   &   &   &   &   &    &   \textbf{(mag)}  & \boldmath{$10^{-15}\rm(erg\,s^{-1}\,cm^{-2})$} \\
\textbf{(1)} & \textbf{(2)} & \textbf{(3)} & \textbf{(4)} & \textbf{(5) }& \textbf{(6)} & \textbf{(7)} & \textbf{(8)} & \textbf{(9)} &\textbf{ (10)} & \textbf{(11) }\\
	\midrule

78 & 7:36:24.820 & +65:36:02.233 & 0.369 &  2.837 $\pm$ 0.205 & 3.446 $\pm$ 0.705 & 1.559 $\pm$ 0.109 & 0.432 $\pm$ 0.054 & 0.980 $\pm$ 0.097 & \nodata  & 5.8  $\pm$  0.4\\
79 & 7:36:27.838 & +65:35:28.437 & 0.389 &  2.788 $\pm$ 0.038 & 3.001 $\pm$ 0.284 & 0.392 $\pm$ 0.010 & 0.475 $\pm$ 0.009 & 0.528 $\pm$ 0.019 & \nodata  & 27.4  $\pm$  0.3\\
80 & 7:36:39.408 & +65:37:10.680 & 0.147 &  2.860 $\pm$ 0.348 & 3.802 $\pm$ 0.738 & 1.278 $\pm$ 0.176 & 0.511 $\pm$ 0.063 & 0.482 $\pm$ 0.055 & 0.613 $\pm$ 0.090  & 16.5  $\pm$  1.7\\
81 & 7:36:45.748 & +65:36:57.650 & 0.102 &  2.837 $\pm$ 0.029 & 1.868 $\pm$ 0.036 & 0.988 $\pm$ 0.011 & 0.438 $\pm$ 0.005 & 0.328 $\pm$ 0.006 & \nodata  & 110.0  $\pm$  1.1\\
82 & 7:36:49.018 & +65:36:52.259 & 0.097 &  2.860 $\pm$ 0.088 & 2.670 $\pm$ 0.179 & 1.092 $\pm$ 0.039 & 0.531 $\pm$ 0.017 & 0.487 $\pm$ 0.017 & 0.948 $\pm$ 0.023  & 75.5  $\pm$  1.9\\
83 & 7:36:51.827 & +65:36:45.214 & 0.097 &  2.822 $\pm$ 0.015 & 1.891 $\pm$ 0.026 & 1.295 $\pm$ 0.008 & 0.428 $\pm$ 0.003 & 0.366 $\pm$ 0.008 & \nodata  & 102.0  $\pm$  0.5\\
84 & 7:36:59.802 & +65:36:33.455 & 0.159 &  2.860 $\pm$ 0.554 & 2.063 $\pm$ 0.633 & 0.755 $\pm$ 0.167 & 0.576 $\pm$ 0.113 & 0.989 $\pm$ 0.178 & 0.708 $\pm$ 0.144  & 12.1  $\pm$  1.9\\
    \bottomrule
\end{tabularx}

\begin{adjustwidth}{+\extralength}{0cm}
\noindent{\footnotesize{Note. Columns: 
    (1) Object number of the \HII~region;
    (2)--(3) object coordinate in J2000; 
    (4) galactocentric distance normalized by $R_{25}$; 
    (5)--(9) {emission-line fluxes normalized by H$\beta$;}
    (10) dust attenuation at the $V$ band; 
    (11) the fluxes of \Hb in units of $10^{-15}$\,erg\,s$^{-1}$\,cm$^{-2}$.}}
\end{adjustwidth}

	\label{table:line_ratio}%
\end{table}%
\finishlandscape

\subsection{Ancillary Data from~the Literature}

\textls[-15]{Complementary with the observed spectra, we additionally compiled a dataset of spectral emission lines for 104 \HII~regions in NGC\,2403 from the literature, all of which have been corrected for dust attenuation.~Specifically, the~supplementary sample contained 6~\HII~regions collected from \citet{McCall1985}, 5 \HII~regions collected from \mbox{\citet{Fierro1986}}, 12 \HII~regions collected from \citet{Garnett1997}, 9 \HII~regions collected from \mbox{\citet{Bresolin1999}}, 3 \HII~regions collected from \citet{Garnett1999}, 17 \HII~regions collected from \citet{van1998}, 7 \HII~regions collected from \citet{Berg2013}, 12 \HII~regions collected from \citet{Mao2018}, and~33 \HII~regions collected from \citet{Rogers2021}.  The~spatial locations for these \HII~regions along with the observed ones are all shown in Figure~\ref{fig:Hii_slit}.}

\begin{figure}[H]
	
	\includegraphics[width=0.56\linewidth]{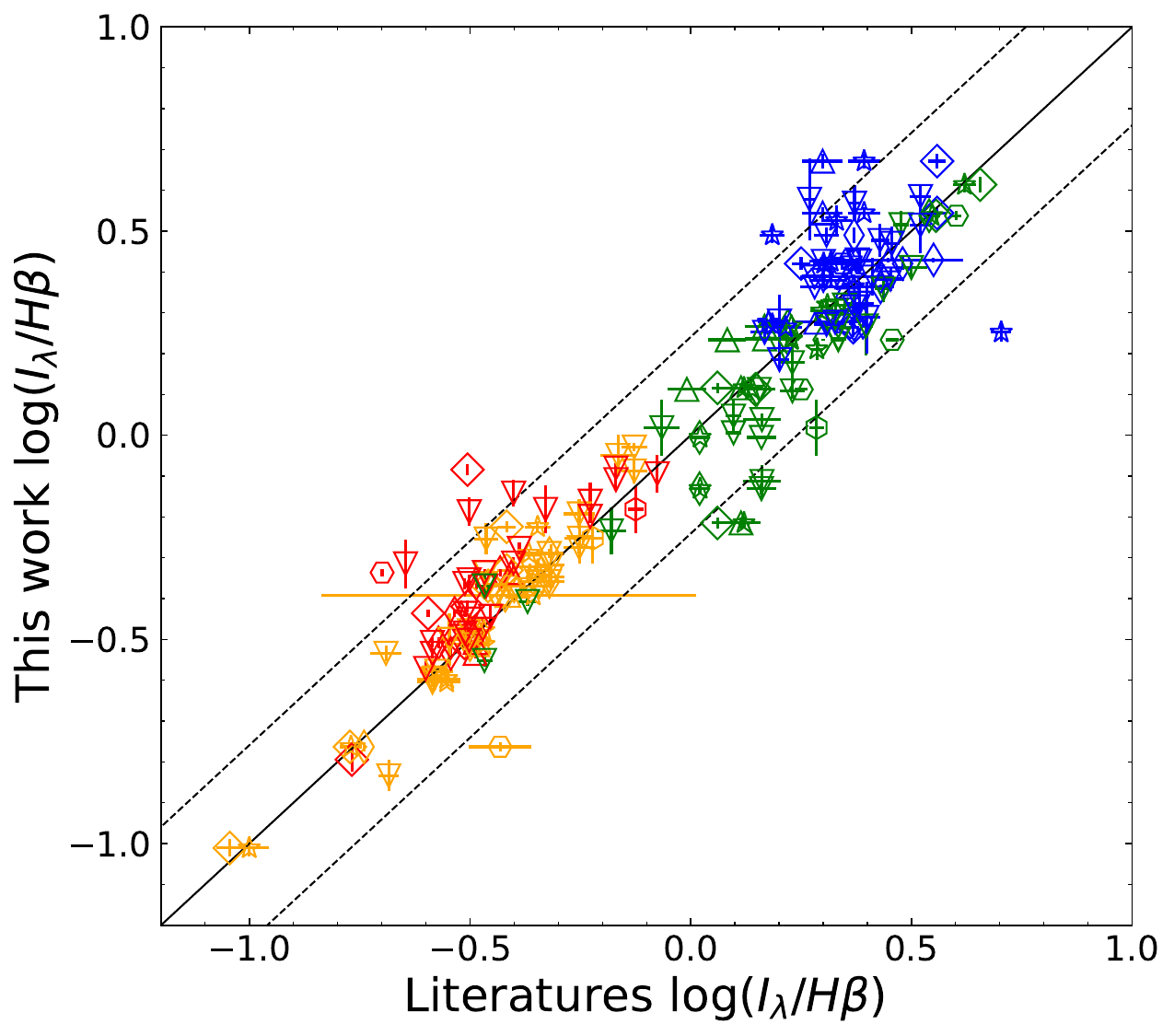}
	\caption{Comparison between the emission-line fluxes for the same \HII~regions measured in this work and those quoted from other studies, including \citet{McCall1985}, \citet{Fierro1986}, \mbox{\citet{Garnett1997}}, \citet{Bresolin1999}, \citet{Garnett1999}, \citet{Mao2018}, and~\citet{Rogers2021}.  The~emission lines include \OII\,$\lambda$3727 (blue), \OIII\,$\lambda\lambda$4959,5007 (green), \NII\,$\lambda$6584 (yellow), and~\SII\,$\lambda\lambda$6717,6731 (red), and they are normalized by H$\beta$. {The data points are marked using the same symbols as in Figure~\ref{table:slit}.}  The line of unity is represented as the black solid line.  The~$\pm 2\sigma$ ($\pm 0.24$ dex, i.e.,~the standard deviation of the data) range are enveloped by the black dotted~lines.}
	\label{fig:line_comparison}
\end{figure}

For the common \HII~regions observed in this work and collected from other studies, we compared the emission-line fluxes, and the result is shown in Figure~\ref{fig:line_comparison}.~Most of the data points were enveloped within the $\pm 2\sigma$ ($\pm 0.24$\,dex) demarcation, which verified the consistency of the data.  Differences between the apertures in size and position for extracting the spectra for the same \HII~regions in the different studies were considered to be a general source of the deviation.  Specifically, a~few of $\OII\lambda3727$ and $\SII\lambda\lambda6717,6731$~lines deviated to a relatively large extent, which was ascribed to influences of different noise levels at the blue end of the spectra on $\OII\lambda3727$ and different deblending qualities for $\SII\lambda\lambda6717,6731$~doublets, respectively.

\unskip

\section{Photoionization~Model}\label{sec:Model}

We adopted model grids developed by \citet{Levesque2010}, which combine the STARBURST99 stellar population synthesis model (\citet{Leitherer1999, Vazquez2005}) with the MAPPINGS III photoionization model (\citet{Sutherland1993ApJS...88..253S}, \mbox{\citet{Allen2008}}).  In~this work, ionizing stellar populations were modeled by assuming star formation history of an instantaneous burst in the range 0--9.5\,Myr with the step length 0.5\,Myr in age, 5 metallicity values including $Z$ = 0.001, 0.004, 0.008, 0.02, and~0.04 (corresponding to 0.05, 0.2, 0.4, 1, and~2\,$Z_{\bigodot}$), and~the Geneva “High” mass-loss star evolutionary tracks (\citet{Schaller1992, Schaerer1993_1, Charbonnel1993, Schaerer1993_2}; recommended in \citet{Levesque2010}).  

Ionized gaseous nebulae were configured in a parallel-plane geometry.  The~ionization parameter $q$ at the inner surface of the nebulae was set to spanning $q=1\times10^7$, $2\times10^7$, $4\times10^7$, $8\times10^7$, $1\times10^8$, $2\times10^8$, $4\times10^8~\mathrm{cm~s^{-1}}$. The~gas-phase metallicity matched with that of the model stellar populations.  The~electron density $n_\mathrm{e} = 100\,\mathrm{cm^{-3}}$ was chosen in accordance with the low-density limit for \HII~regions in NGC\,2403 (\citet{Garnett1997,Berg2013,Rogers2021}).

\section{Results} \label{sec:Result}

Strong emission lines in optical spectra for ionized gaseous nebulae encode a wealth of information and serve as effective diagnostics for a series of physical parameters.  In~this section, we present results of diagnoses from a series of strong-line indices for the \HII~regions in NGC\,2403 and in turn inspect influences of underlying parameters on the emission~lines.
  
\subsection{BPT~Diagram}

\begin{figure}[H]
	
	\includegraphics[width=0.56\linewidth]{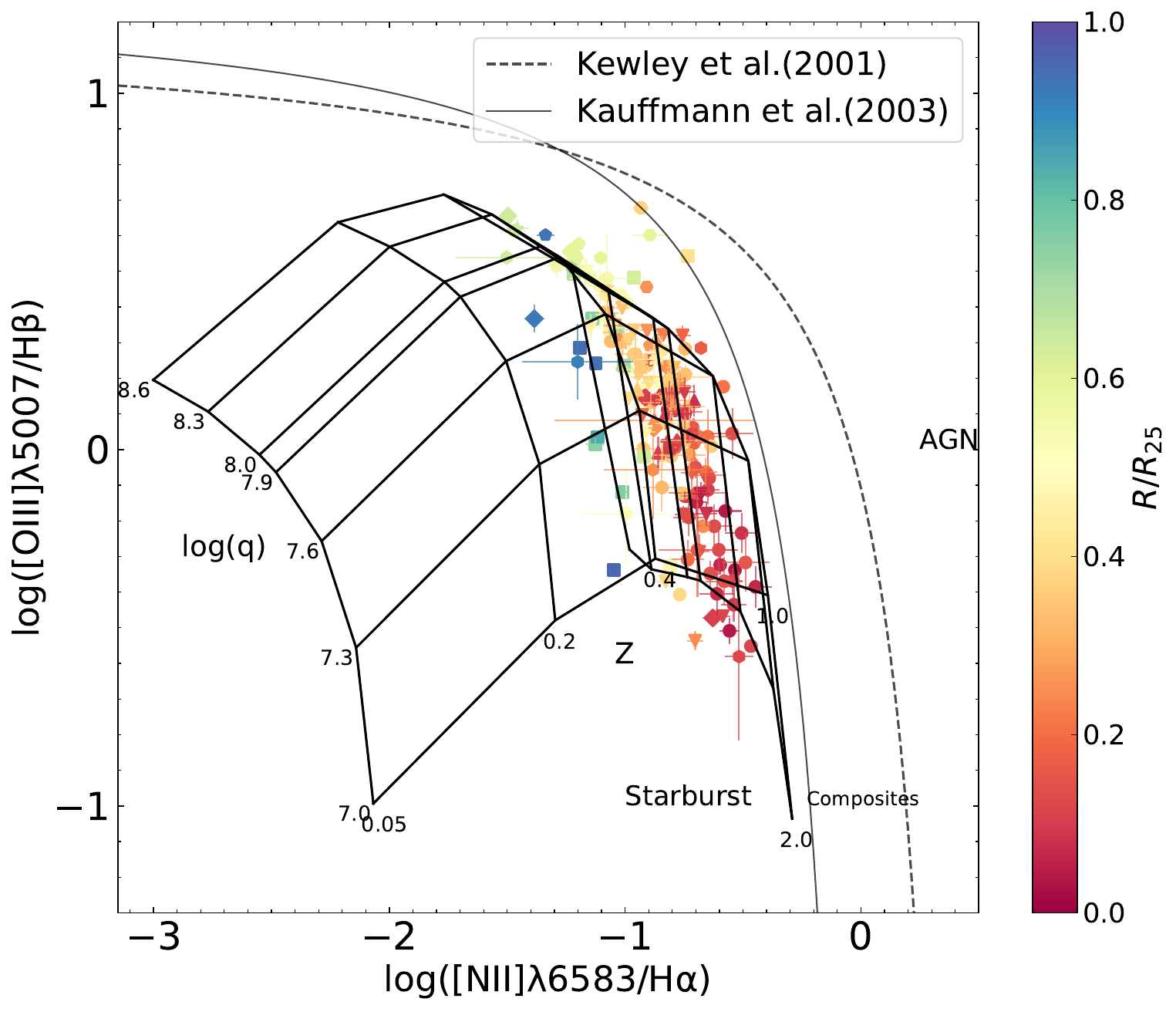}
	\caption{$\log(\OIII\lambda 5007/\Hb)$ as a function of $\log (\NII\lambda 6583/\Hb)$ (the BPT diagram) for the \HII~regions in NGC\,2403 with the \citet{Levesque2010} model grid (the constant age 0\,Myr, variations in the metallicity Z (in units of $Z_{\bigodot}$), and the ionization parameter q) superimposed.  {The data points are marked using the same symbols as in Figure~\ref{table:slit}} and color-coded by the galactocentric distance (de-projected) in units of $R_{25}$.  The~criteria for classifying ionizing sources (stars and AGNs) are quoted from \citet[][]{Kewley_2001} (black dotted line) and \citet[][]{Kauffmann_2003} (black solid line).}
	\label{fig:BPT_dig}
\end{figure}

The diagnostic diagram of the indices $\OIII\lambda 5007/\Hb$ versus (vs.) $\NII\lambda 6583/\Hb$ proposed by Baldwin, Phillips, and~Terlevich (denoted as BPT, \citet{BPT1981}) is a powerful tool for classifying emission-line objects into active galactic nuclei (AGNs), low-ionization-emission regions (LINERs), or~star-forming regions (\citet{Kewley_2001,Kauffmann_2003}).  Figure~\ref{fig:BPT_dig} is the BPT diagram for the \HII~regions in NGC\,2403, with~the 0.0\,Myr model grid as well as the \citet{Kewley_2001} and \citet{Kauffmann_2003} demarcation curves for the classification superimposed.  {Almost all of the points are consistent with the \citet{Kauffmann_2003} curve, and~only one falls between the \citet{Kauffmann_2003} and \mbox{\citet{Kewley_2001}} curves, indicating that nearly all samples originate from star formation activity.} By comparing the color-coded data points with the model grid,  the~data locus in the BPT diagram reflects a sequence of metallicity or/and the ionization parameter varying with the galactocentric distance and~suggests a correlation of the two indices with both metallicity and the ionization~parameter.

\unskip

\subsection{Metallicity~Estimates}

The oxygen abundance for the \HII~regions in NGC\,2403 was estimated for assessing the metallicity.  There are a number of recipes for obtaining the oxygen abundance but considerable discrepancies among them, leaving an important challenge in metallicity studies (\mbox{\citet{Kewley2008, Mao2018, Kewley2019, maiolino2019}}).  In this work, we adopted four widely used empirical diagnostics for estimating oxygen abundance.  They are described as~follows.

$R23$ (\citet{Pilyugin_Thuan2005ApJ...631..231P}):\endnote{The $R23$ method has a double-value effect.  We choose the higher branch in this work, because~the \HII~regions in NGC\,2403 are generally metal-rich according to our initial assessment.}
\begin{equation}\label{eq:R23}
12+\log(O/H)_{high} = \frac{R23 + 726.1 + 842.2P + 337.5P^2}{85.96+ 82.76P+43.98P^2+1.793R23}  ,
\end{equation}
where $R23 \equiv \log\frac{\OII\lambda3727+\OIII\lambda\lambda4959,5007}{\Hb}$, $P=\frac{\OIII\lambda\lambda4959,5007}{\OIII\lambda\lambda4959,5007+\OII\lambda3727}$;

$O3N2$ (\citet{PP2004}):
\begin{equation}\label{O3N2}
12+\log(O/H) = 8.73 - 0.32O3N2  ,
\end{equation}
where $O3N2 \equiv \log\frac{\OIII\lambda5007/\Hb}{\NII\lambda6583/\Ha}$;

$N2$ (\citet{PP2004}):
\begin{equation}\label{N2}
12+\log(O/H) = 9.37+2.03 \times N2+1.26\times N2^2+0.32\times N2^3  ,
\end{equation}
where $N2 \equiv \log\frac{\NII\lambda6583}{\Ha}$;

$N2O2$ (\citet{Bresolin2007}):
\begin{equation}\label{eq:N2O2}
12+\log(O/H) = 8.66 + 0.36 \times N2O2 - 0.17\times N2O2^2  ,
\end{equation}
where $N2O2 \equiv \log\frac{\NII\lambda6583}{\OII\lambda3727}$.

\begin{figure}[H]
	
	\includegraphics[width=0.97\linewidth]{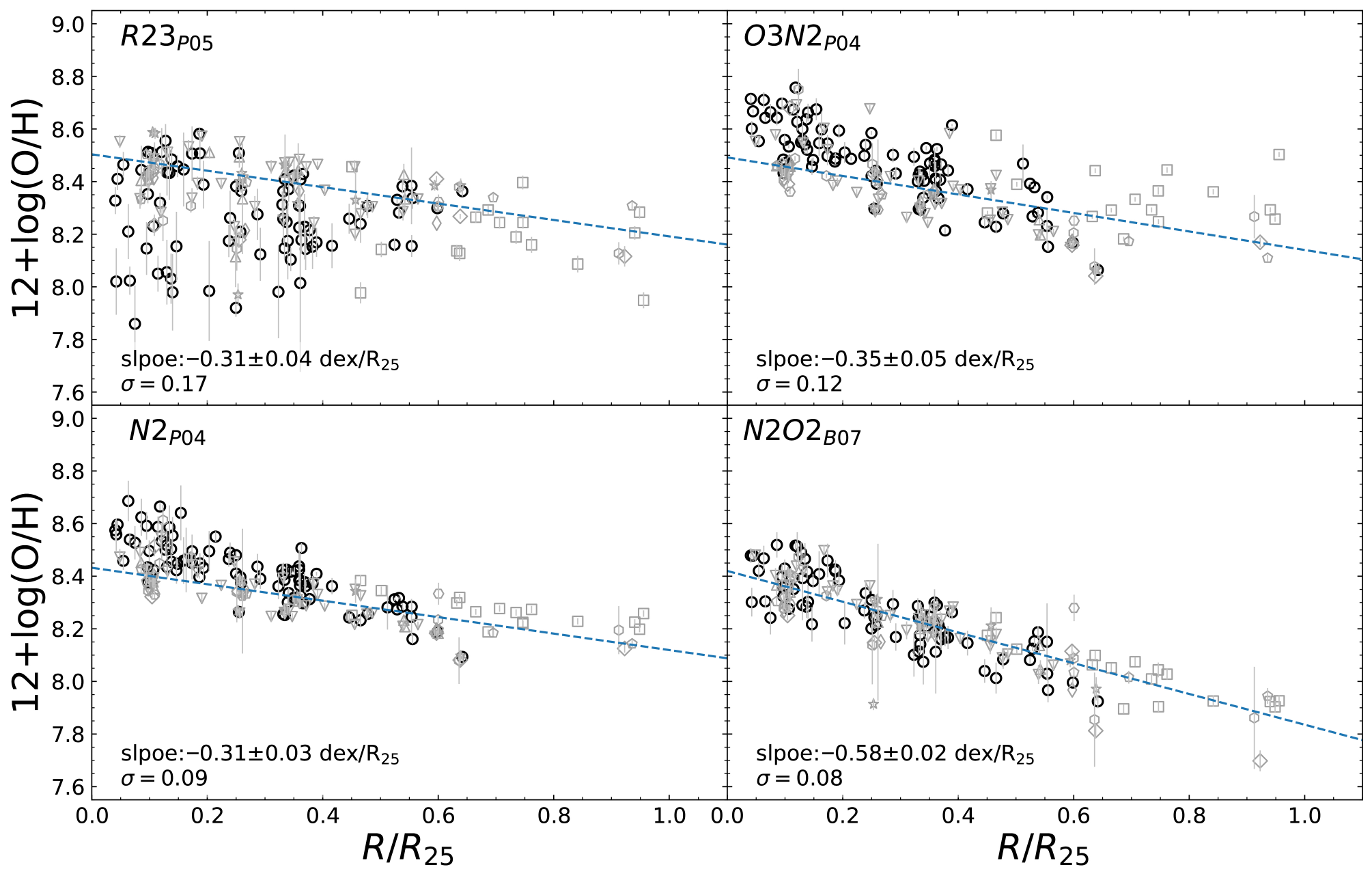}
	\caption{{Radial} 
 distributions of the oxygen abundance derived from the indices $R23$ ((\textbf{top-left panel}), the~calibration is addressed in \citet[][]{Pilyugin_Thuan2005ApJ...631..231P} and denoted as $R23_\mathrm{P05}$), $O3N2$ ((\textbf{top-right panel}), the~calibration is addressed in \citet[][]{PP2004} and denoted as $O3N2_\mathrm{P04}$), $N2$ ((\textbf{bottom-left panel}), the~calibration is addressed in \citet[][]{PP2004} and denoted as $N2_\mathrm{P04}$), and~$N2O2$ ((\textbf{bottom-right panel}), the~calibration is addressed in \citet[][]{Bresolin2007} and denoted as $N2O2_\mathrm{B07}$) for the \HII~regions (represented by the same symbols as assigned in Figure~\ref{fig:Hii_slit}) in NGC\,2403 with the weighted linear best-fit curves (cyan dashed lines) superimposed. {The fitting parameters are compiled in Table~\ref{table: metallicity}.}
    }
	\label{fig:metallicity}
\end{figure}

From the above diagnostics, we estimated the oxygen abundances for the \HII~regions in NGC\,2403, whose radial profiles are displayed in Figure~\ref{fig:metallicity}, with~the best-fit linear curves superimposed. {The uncertainty of the slope was quantified as a standard deviation derived from bootstrapping the sample 3000 times, and the corresponding fit parameters are listed in Table~\ref{table: metallicity}.}  A negative gradient is obviously seen from each of the profiles, consistent with the inside-out scenario of galaxy evolution.  However, differences in gradient slope and data dispersion between these radial profiles are evident and imply greater complexity for the emission-line indices, which inevitably affects and biases relevant studies based on metallicity gradients derived from these~recipes.

\begin{table}[H]
	
	\caption{Fitting parameters of radial metallicity~distribution.}\label{table: metallicity}
	\newcolumntype{C}{>{\centering\arraybackslash}X}
\begin{tabularx}{\textwidth}{CCCC}
	\toprule
	
    \textbf{Indicator} & \textbf{Gradient} & \boldmath{$\sigma$} & \boldmath{$R^2$}\\
          &\textbf{(\boldmath{$\rm dex/R_{25}$})}&\textbf{(\boldmath{$\rm dex$})}&\\
    \textbf{(1)} & \textbf{(2)} & \textbf{(3) }& \textbf{(4)}\\
        \midrule
        $R23_{P05}$ & $-$0.31 $\pm$ 0.04 & 0.17 & 0.35\\ 
        $O3N2_{P04}$ & $-$0.35 $\pm$ 0.05 & 0.12 & 0.41 \\ 
        $N2_{P04}$ & $-$0.31 $\pm$ 0.03 & 0.09 & 0.58\\ 
        $N2O2_{B07}$ & $-$0.58 $\pm$ 0.02 & 0.08& 0.81\\ 
	\bottomrule
	\end{tabularx}
	
	\noindent{\footnotesize{{Notes.} Columns: (1) Metallicity indicator; (2)–(4) slope of the radial metallicity gradient, dispersion, and~coefficient of determination of the fit, as~presented in Figure~\ref{fig:metallicity}.}}

\end{table}%

\unskip

\subsection{Index--Index~Diagrams}\label{sec:diag}

\textls[-15]{In order to inspect the emission-line indices in more depth, we carried out a series of diagnoses by diagramming relations of the indices for the data in combination with the model grids.  These diagnoses result in a further understanding of not only underlying parameters in the observed features but also more properties for the \HII~regions in NGC\,2403.  In~this section, the~emission-line indices inspected in the diagnoses included the metallicity indicators $R23$, $O3N2$, $N2$, and~$N2O2$. As~a widely used proxy of the ionization parameter, $O3O2$ ($\equiv \OIII\lambda\lambda4959,5007/\OII\lambda3727$) was also employed in our work for specially tracing ionization states.\endnote{In observational studies, a~ratio between two emission lines for a certain ion at different ionization states is usually taken as a proxy of the ionization parameter (e.g., \citet{Rela2010, Mao2018, Kewley2019}).~The~emission-line indices $\OIII\lambda\lambda4959,5007/\OII\lambda3727$ (denoted as $O3O2$) and $\SIII \lambda\lambda 9069,9531/\SII \lambda\lambda 6717,6731$ (denoted as $S3S2$ \mbox{\citet{Dopita2000,KD2002,Morisset2016A&A...594A..37M, Kewley2019, Garner2025ApJ...978...70G}}) are the most widely used ionization tracers.  Owing to the wavelength coverage in our observations, we employed $O3O2$ in the diagnoses. }}

\subsubsection{Diagnosis at Fixed~Age}\label{sec:diag_0Myr}

\begin{figure}[H]
	
	\includegraphics[width=\linewidth]{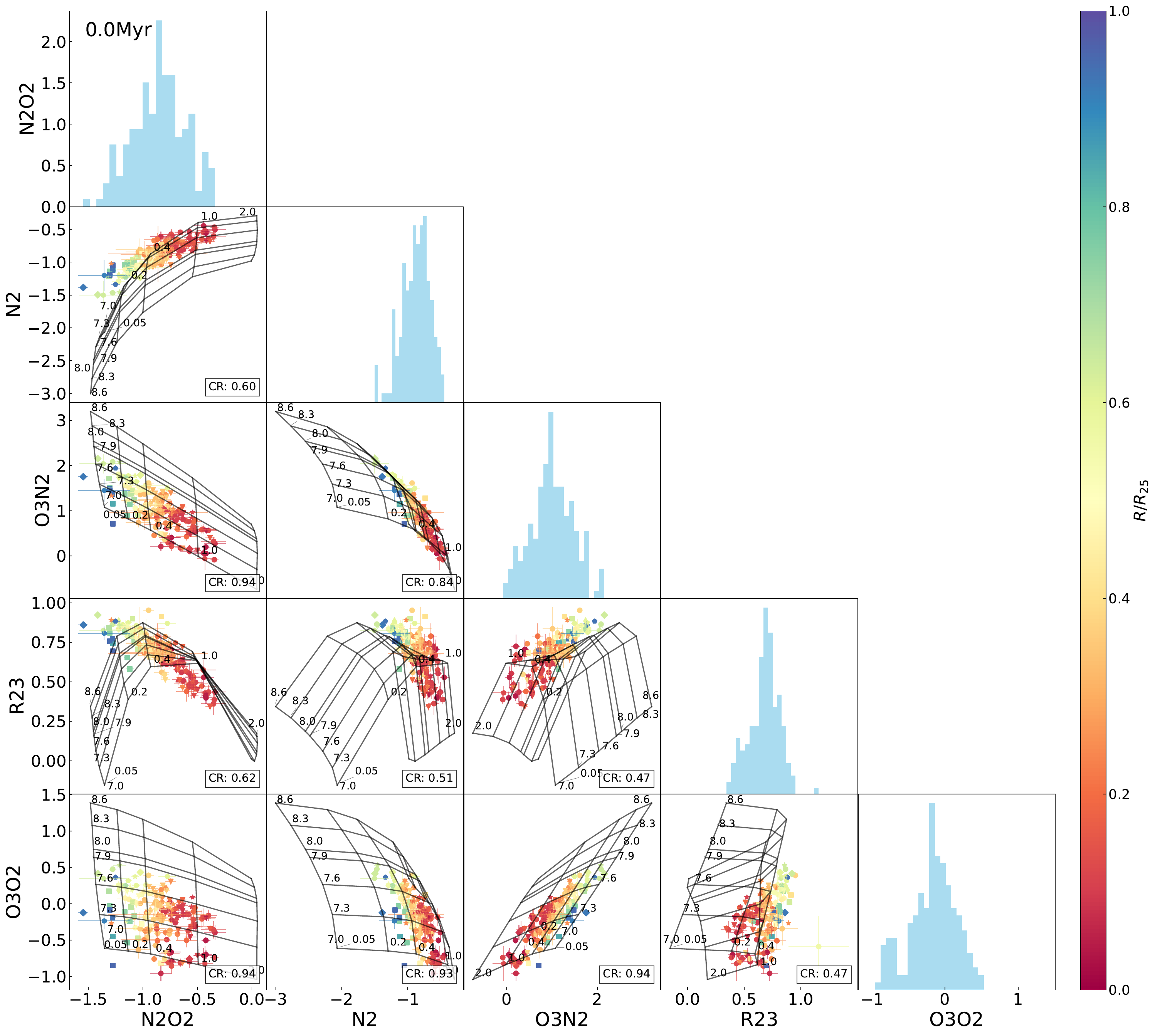}
	\caption{{Interrelations} 
 between every two indices among $R23$, $O3N2$, $N2$, $N2O2$, $O3O2$ for the \HII~regions in NGC\,2403, with~the \citet{Levesque2010} model grid (the constant age of 0\,Myr, variations in metallicity Z (in units of $Z_{\bigodot}$) and the ionization parameter q) superimposed. {The data points are marked using the same symbols as in Figure~\ref{table:slit}} and color-coded the same as in Figure~\ref{fig:BPT_dig}.  The~rate of the data covered by the model grid (defining the number ratio of the data points falling within the grid to the total of the data points, denoted as the coverage rate or ``CR'') is presented in the lower-right corner in each of the panels. Histograms of the indices are also~shown.}
	\label{fig:index_grid_0Myr}
\end{figure}

Figure~\ref{fig:index_grid_0Myr} shows the interrelations between $R23$, $O3N2$, $N2$, $N2O2$, and~$O3O2$, with~the model grid reflecting the initial situation (i.e., the~0\,Myr age) superimposed in each of the panels.  {As illustrated in these diagrams, among~the four metallicity indicators, the~correlation with $O3O2$ is strong for $O3N2$ and $N2$ (Pearson coefficients {$\rho = 0.88$ and $-0.72$}, respectively), moderate for $R23$ {($\rho = 0.45$)}, and~weak for $N2O2$ {($\rho = -0.39$)}.}  Different levels of degeneracy of metallicity and the ionization parameter in the different indices are disclosed by the model grids.  Specifically, for~constant values of metallicity, $O3N2$ increases monotonically with the ionization parameter, whereas $N2$ decreases at higher ionization states with a slight curvature; dependence of $R23$ on the ionization parameter appears complicated, with~two opposite trends to the low- and high-metallicity ends, similar to the so-called double-value effect of $R23$ on its estimates of the oxygen abundance; by contrast, $N2O2$ stays robust against variations in the ionization parameter, as~manifested by the extensive model grid and the widespread data distribution in the $O3O2$-$N2O2$ diagram in Figure~\ref{fig:index_grid_0Myr}.  This result verifies the non-negligible sensitivities of $O3N2$ and $N2$ to the ionization parameter, as~well as the reliability of $N2O2$ for estimating metallicity, particularly at variable ionization states.  The~sensitivity of $R23$ to the ionization parameter was inherently weaker than that of $O3N2$ and $N2$; furthermore, the correction for the ionization influence was taken into account in the $R23$ calibration (\mbox{\citet{Pilyugin_Thuan2005ApJ...631..231P}}).  

Nonetheless, the~data points from the outer part of the galaxy ($r > 0.2 R_{25}$) lie in the conjunction zone of the higher and lower branches of the $R23$ calibration and thus appear with dispersion in the $R23$-based estimates of metallicity.  For~the \HII~regions located in the inner area of the galaxy ($r \lesssim 0.2 R_{25}$), even larger dispersion emerges in the $R23$-based estimates of metallicity (see the top-left panel of Figure~\ref{fig:metallicity}).  This is likely due to an inherent problem in the \citet{Pilyugin_Thuan2005ApJ...631..231P} calibration.  We found these data points in possession of lower excitation levels (quantified by the excitation parameter $P$ as defined in Equation~(\ref{eq:R23})).  As~pointed out in \citet{Moustakas2010ApJS..190..233M}, the~R23 calibration lacks sufficient low-excitation ($P \lesssim 0.2$) data, which prevents its good performance in such a regime and thereby induces the dispersion in estimates of metallicity.\endnote{Although the ionization parameter was taken into account, the~sensitivity of the \citet{Pilyugin_Thuan2005ApJ...631..231P} calibration to the ionization parameter was still found in our further analysis ({Li~et~al.,} 
 in~preparation), which points to the complexity of the $R23$ index as a reliable or questionable estimator of metallicity.}

{The $N2O2$ metallicity diagnostic is based on the relation of $N/O$–$O/H$ at high metallicity, where secondary nitrogen production leads to a correlation between $N/O$ and $O/H$ {(\citet{KD2002,maiolino2019})}. For~NGC\,2403, \citet{Berg2013} found that within $R \sim 0.93\,R_{25}$ (our estimate), $\log(N/O)$ and $12+\log(O/H)$ exhibited a tight linear correlation, without~the plateau characteristic of primary nitrogen production. Together with its insensitivity to the ionization parameter and immunity to the double-valued degeneracy, this supports the robustness of $N2O2$ as a metallicity indicator.}

As a consequence, among~the four panels in Figure~\ref{fig:metallicity}, the~{metallicity estimated from $N2O2$} shows the most reliable radial metallicity gradient for NGC\,2403, {with a slope of $-0.58 \pm 0.02$\,{dex\,R$_{25}^{-1}$}, a~dispersion of $\sigma = 0.08$\,dex, and~a coefficient of determination of $R^2 = 0.81$}.

Comparison of the data distribution with the model grid in each of the index--index diagrams in Figure~\ref{fig:index_grid_0Myr} manifests different extents of coverage of the model to the data points. Specifically, there are approximately 93--94\% of the data points covered by the model grid in each of the $O3N2$-$N2O2$, $O3O2$-$N2$, and~$O3O2$-$O3N2$ diagrams, about 84\% in the $O3N2$–$N2$ diagram, and~around 60\% or even less in the $N2O2$-$N2$ and $R23$-relative diagrams.  The~model appears to characterize the data well in the $O3N2$-$N2O2$, $O3O2$-$N2$, $O3O2$-$O3N2$, and~$O3N2$–$N2$ diagrams but~not in the others.  In~each of the $N2O2$-$N2$ and $R23$-relative diagrams, the~data locus fails to follow the bending regime of the model grid but~keeps a more linear trend in the plane.  As~a result, some of the data points exceed the model coverage, and~they are the \HII~regions located in the outskirts of NGC\,2403.

\subsubsection{Diagnosis with Age~Evolution}\label{sec:diag_age}

\begin{figure}[H]
\includegraphics[width=\linewidth]{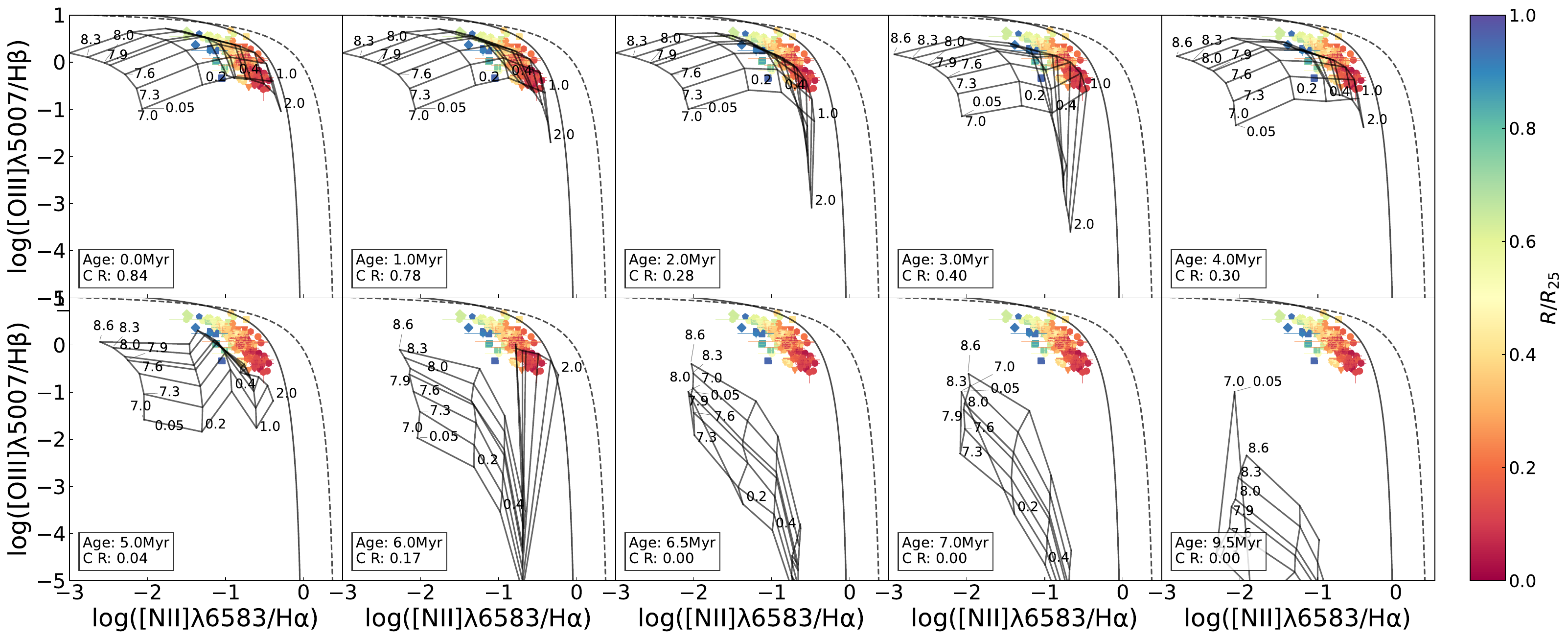}
\caption{{BPT} 
 diagrams for the \HII~regions in NGC\,2403 with the \citet{Levesque2010} model grids (constant age, variations in metallicity Z (in units of $Z_{\bigodot}$) and the ionization parameter q in each of the panels) superimposed. The~data points and the~dotted and solid lines have the same meaning in all of the panels as in Figure~\ref{fig:BPT_dig}).  In~the different panels, the~model grids vary with the age ranging from 0 to 9.5\,Myr {(This only shows part of the age grid as a representative.)}  {The age and the CR (the same definition as in Figure~\ref{fig:index_grid_0Myr}) are listed in the lower-right corner in each of panels.} {The data points are marked using the same symbols as in Figure~\ref{table:slit}} and color-coded the same as in Figure~\ref{fig:BPT_dig}.
}
\label{fig:BPT_age}
\end{figure}

\begin{figure}[H]
\includegraphics[width=\linewidth]{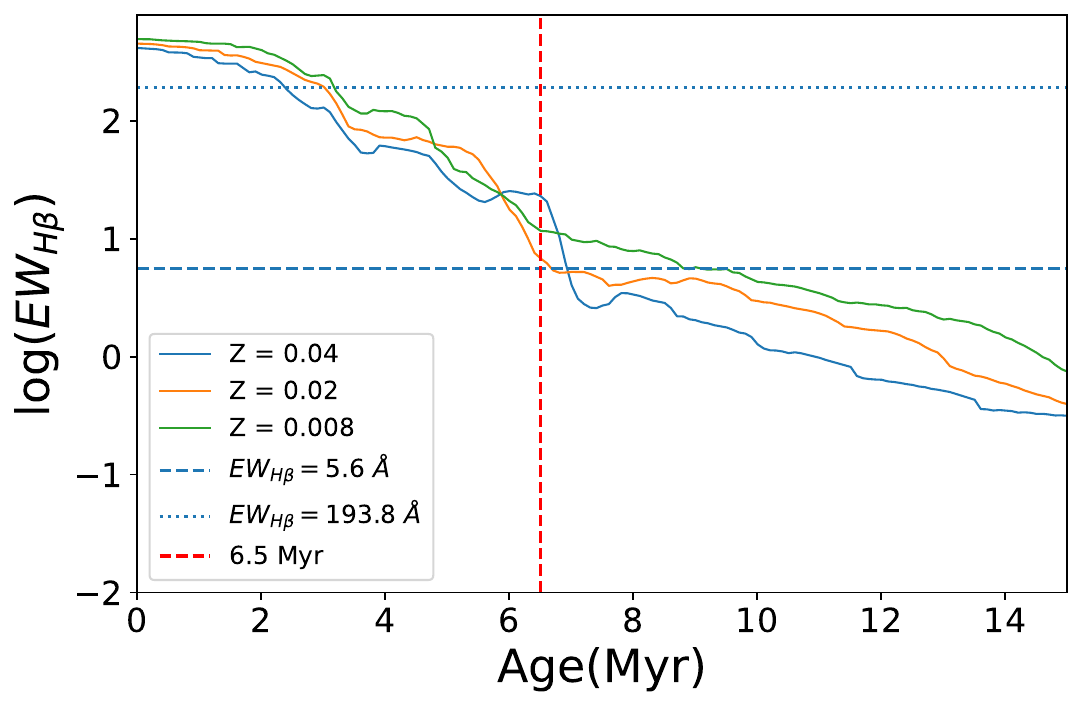}
\caption{Constraints on the stellar population age derived from the equivalent width of the {$\Hb$} emission line ({$EW_{\Hb}$}). The~model curves show {$EW_{\Hb}$}–age relations for instantaneous bursts at different metallicities from the STARBURST99 model {(\citet{Leitherer1999})} 
 using the publicly available 1999 dataset downloaded from the official website. Horizontal dotted and dashed lines mark the observed maximum and minimum $EW_{\Hb}$ values of the emission lines across the galaxy. The~red vertical line indicates the 6.5 Myr upper-age limit inferred from the BPT diagram coverage. The~minimum {$EW_{\Hb}$} value is found in the central, higher-metallicity region, consistent with an age of $\sim$7 Myr (Z = 0.02).}
\label{fig:ew_age}
\end{figure}

\begin{figure}[H]
\includegraphics[width=\linewidth]{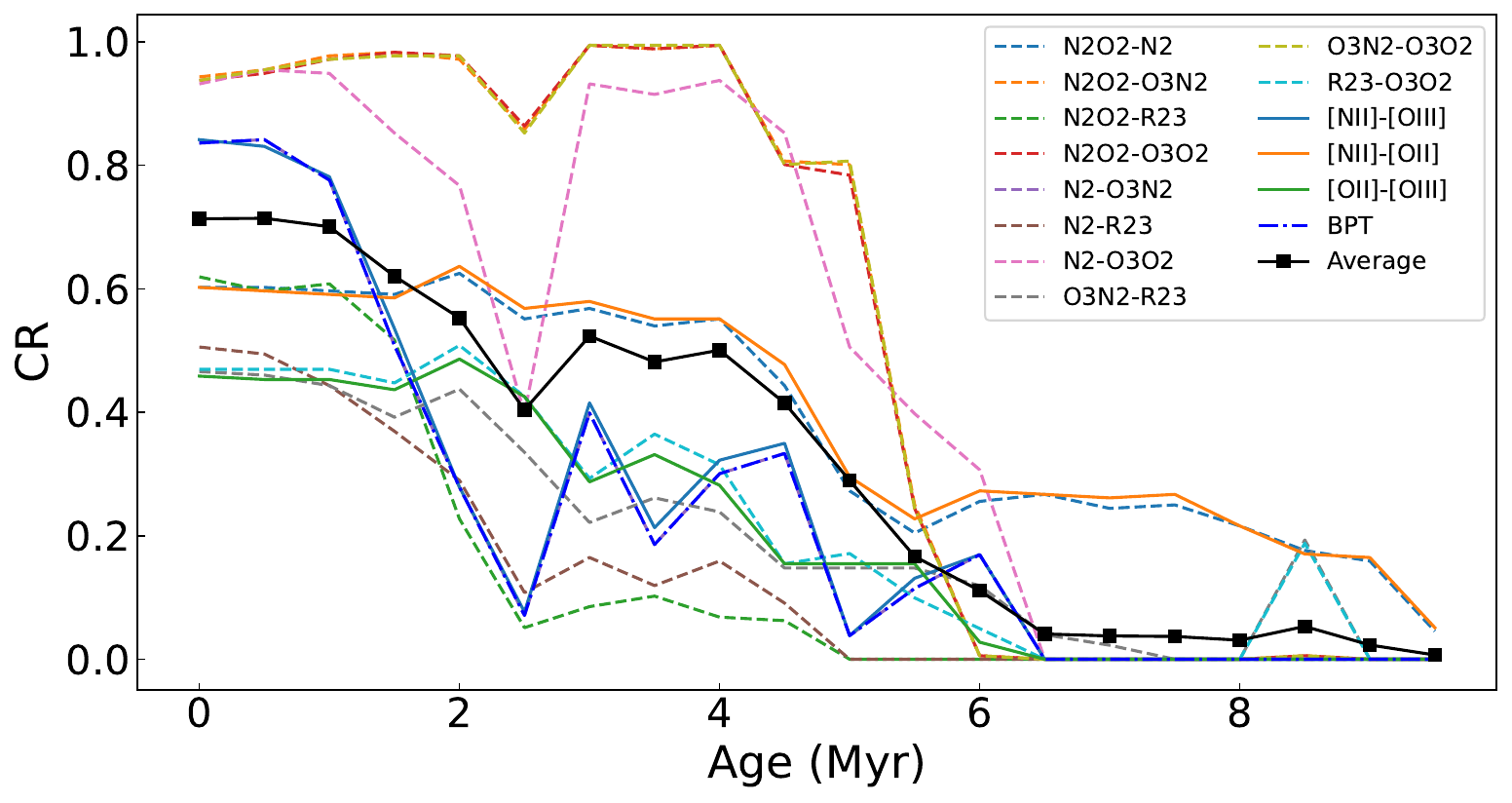}
\caption{The {CR (the same definition as in Figure~\ref{fig:index_grid_0Myr})} of the model grid as a function of the age ranging from 0 to 9.5\,Myr, for~the diagrams in Figures~\ref{fig:BPT_dig}, \ref{fig:index_grid_0Myr} and \ref{fig:emission_line_grid_0Myr}.  The~dashed lines are color-coded by specific index--index relations with the legend displayed in the top-right corner.  The~black filled boxes connected by the black solid line represent the average evolution of the {CR} among all of the~diagrams.
}
\label{fig:CR}

\end{figure}

\begin{figure}[H]
\includegraphics[width=0.97\linewidth]{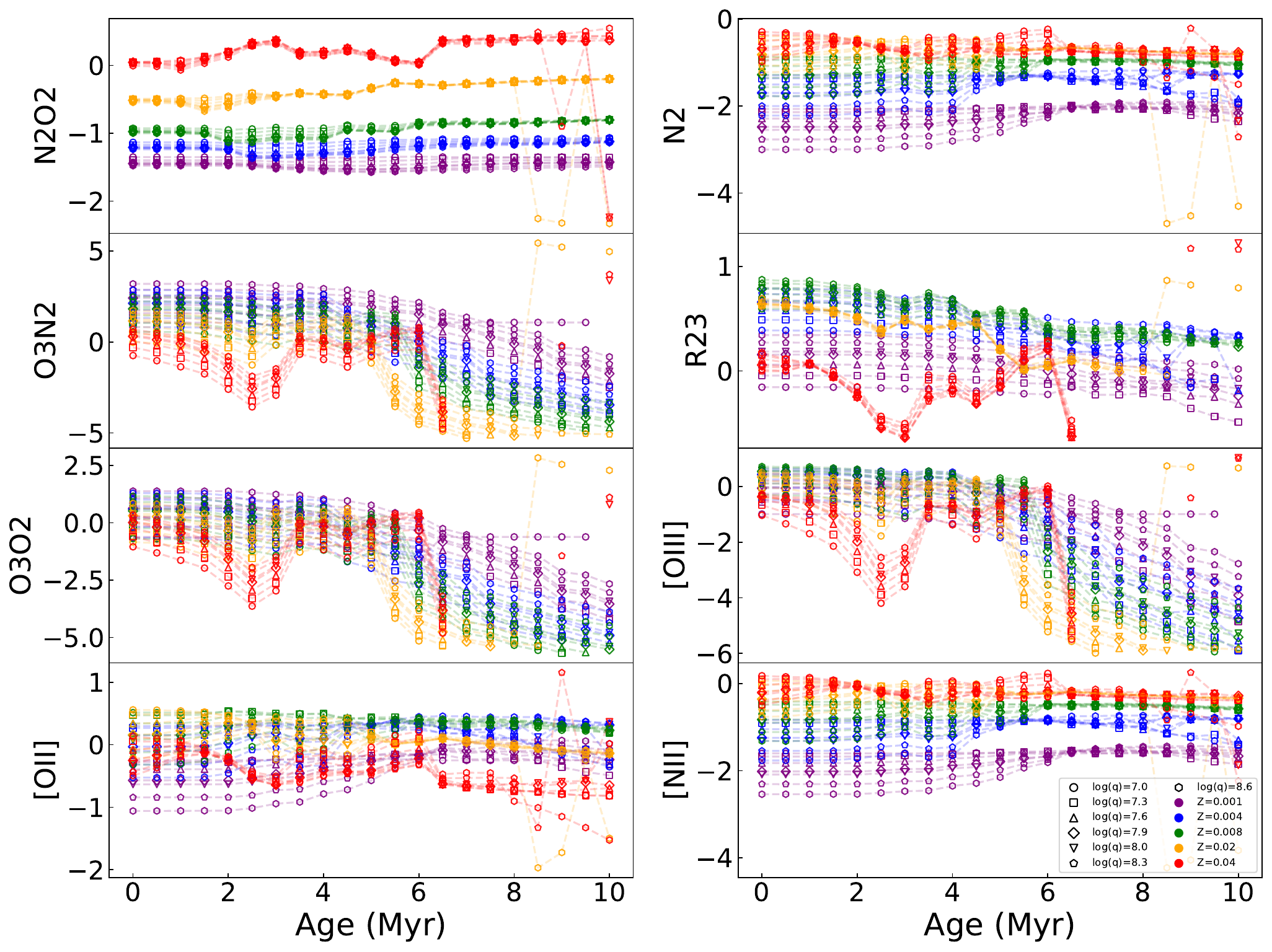}
\caption{{Spectral index or H$\beta$-normalized emission lines of \OIII$\lambda5007$, \NII$\lambda6583$, and~\OII$\lambda3727$ (hereafter referred to as \OIII, \NII, and~\OII, respectively)} as a function of age reproduced by the \citet{Levesque2010} model.  In~each of the panels, the~model products are color-coded by metallicity and symbol-coded by the ionization parameter, with~the legend listed in the lower-left corner in the bottom-right panel.
}
\label{fig:index_age}
\end{figure}

The model grids in Figures~\ref{fig:BPT_dig} and \ref{fig:index_grid_0Myr} were established at the fixed age of 0\,Myr (in an initial situation) for the purpose of eliminating effects of age but specially focusing on behaviors of metallicity and the ionization parameter.  As~a matter of fact, emission lines are supposed to tightly relate with the age of ionizing stellar populations, due to a decrease in the number of ionizing photons when stellar populations age.  With~the diagrammatic diagnosis independent of age worked out above, we conducted modeling of the spectral indices by varying the age (range of 0--9.5\,Myr) in this section, to~inspect the age evolution of the index--index relationships, whose results are shown in Figures~\ref{fig:BPT_age}--\ref{fig:index_age} and Appendix \ref{sec:App_index_age}. {The coverage rates for all diagnostic diagrams are summarized in Table~\ref{table: CR}.}

\startlandscape

\begin{table}[H]

\caption{{Coverage} 
 rates of diagnostic~diagram.}\label{table: CR}
\scriptsize

\newcolumntype{C}{>{\centering\arraybackslash}X}
\begin{tabularx}{\textwidth}{Cccccccccccccccc}

\toprule
\textbf{AGE} & \textbf{BPT}  & \boldmath{$N2O2-N2$} & \boldmath{$N2O2-O3N2$} & \boldmath{$N2O2-R23$} & \boldmath{$N2O2-O3O2$} & \boldmath{$N2-O3N2$} & \boldmath{$N2-R23$} & \boldmath{$N2-O3O2$} & \boldmath{$O3N2-R23$} & \boldmath{$O3N2-O3O2$} & \boldmath{$R23-O3O2$} & \pmb{\NII-\OIII} & \pmb{\NII-\OII} & \pmb{\OII-\OIII} & \textbf{Average} \\ 
\textbf{(Myr)} &   &  &  &  &  &  & &  & &  & &  &  &  & \\
\textbf{(1)} &\textbf{ (2)}  &\textbf{ (3)} & \textbf{(4)} & \textbf{(5)} & \textbf{(6) }& \textbf{(7)} & \textbf{(8)}& \textbf{(9)} & \textbf{(10)}& \textbf{(11)} & \textbf{(12)} & \textbf{(13)} & \textbf{(14)} & \textbf{(15)} & \textbf{(16)} \\
\midrule
0   & 0.84 & 0.60    & 0.94      & 0.62     & 0.94      & 0.84    & 0.51   & 0.93    & 0.47     & 0.94      & 0.47     & 0.84                 & 0.60                & 0.46                 & 0.71    \\
0.5 & 0.84 & 0.60    & 0.95      & 0.60     & 0.95      & 0.84    & 0.49   & 0.95    & 0.46     & 0.95      & 0.47     & 0.83                 & 0.60                & 0.45                 & 0.71    \\
1   & 0.78 & 0.60    & 0.98      & 0.61     & 0.97      & 0.78    & 0.44   & 0.95    & 0.44     & 0.97      & 0.47     & 0.78                 & 0.59                & 0.45                 & 0.70    \\
1.5 & 0.51 & 0.59    & 0.98      & 0.52     & 0.98      & 0.51    & 0.37   & 0.85    & 0.39     & 0.98      & 0.45     & 0.54                 & 0.59                & 0.44                 & 0.62    \\
2   & 0.28 & 0.63    & 0.97      & 0.23     & 0.98      & 0.28    & 0.29   & 0.77    & 0.44     & 0.98      & 0.51     & 0.28                 & 0.64                & 0.49                 & 0.55    \\
2.5 & 0.07 & 0.55    & 0.86      & 0.05     & 0.86      & 0.07    & 0.11   & 0.40    & 0.34     & 0.85      & 0.43     & 0.08                 & 0.57                & 0.43                 & 0.40    \\
3   & 0.40 & 0.57    & 0.99      & 0.09     & 0.99      & 0.40    & 0.16   & 0.93    & 0.22     & 0.99      & 0.29     & 0.42                 & 0.58                & 0.29                 & 0.52    \\
3.5 & 0.19 & 0.54    & 0.99      & 0.10     & 0.99      & 0.19    & 0.12   & 0.91    & 0.26     & 0.99      & 0.36     & 0.21                 & 0.55                & 0.33                 & 0.48    \\
4   & 0.30 & 0.55    & 0.99      & 0.07     & 0.99      & 0.30    & 0.16   & 0.94    & 0.24     & 0.99      & 0.31     & 0.32                 & 0.55                & 0.28                 & 0.50    \\
4.5 & 0.33 & 0.44    & 0.81      & 0.06     & 0.80      & 0.33    & 0.09   & 0.85    & 0.15     & 0.80      & 0.15     & 0.35                 & 0.48                & 0.15                 & 0.41    \\
5   & 0.04 & 0.27    & 0.80      & 0.00     & 0.78      & 0.04    & 0.00   & 0.51    & 0.15     & 0.81      & 0.17     & 0.04                 & 0.30                & 0.15                 & 0.29    \\
5.5 & 0.11 & 0.20    & 0.25      & 0.00     & 0.24      & 0.11    & 0.00   & 0.40    & 0.15     & 0.24      & 0.10     & 0.13                 & 0.23                & 0.15                 & 0.17    \\
6   & 0.17 & 0.26    & 0.01      & 0.00     & 0.01      & 0.17    & 0.00   & 0.31    & 0.12     & 0.01      & 0.05     & 0.17                 & 0.27                & 0.03                 & 0.11    \\
6.5 & 0.00 & 0.27    & 0.00      & 0.00     & 0.00      & 0.00    & 0.00   & 0.00    & 0.04     & 0.00      & 0.00     & 0.00                 & 0.27                & 0.00                 & 0.04    \\
7   & 0.00 & 0.24    & 0.00      & 0.00     & 0.00      & 0.00    & 0.00   & 0.00    & 0.02     & 0.00      & 0.00     & 0.00                 & 0.26                & 0.00                 & 0.04    \\
7.5 & 0.00 & 0.25    & 0.00      & 0.00     & 0.00      & 0.00    & 0.00   & 0.00    & 0.00     & 0.00      & 0.00     & 0.00                 & 0.27                & 0.00                 & 0.04    \\
8   & 0.00 & 0.22    & 0.00      & 0.00     & 0.00      & 0.00    & 0.00   & 0.00    & 0.00     & 0.00      & 0.00     & 0.00                 & 0.22                & 0.00                 & 0.03    \\
8.5 & 0.00 & 0.18    & 0.01      & 0.00     & 0.01      & 0.00    & 0.00   & 0.00    & 0.19     & 0.01      & 0.19     & 0.00                 & 0.17                & 0.00                 & 0.05    \\
9   & 0.00 & 0.16    & 0.00      & 0.00     & 0.00      & 0.00    & 0.00   & 0.00    & 0.00     & 0.00      & 0.00     & 0.00                 & 0.16                & 0.00                 & 0.02    \\
9.5 & 0.00 & 0.05    & 0.00      & 0.00     & 0.00      & 0.00    & 0.00   & 0.00    & 0.00     & 0.00      & 0.00     & 0.00                 & 0.05                & 0.00                 & 0.01    \\ \bottomrule
\end{tabularx}

\begin{adjustwidth}{+\extralength}{0cm}
\noindent{\footnotesize{{Notes.} Columns: (1) Age of stellar population; (2)–(16) CR of diagnostic diagrams.}}
\end{adjustwidth}

\end{table}

\finishlandscape

\vspace{-8pt}

Figure~\ref{fig:BPT_age} shows a modeled evolution of the BPT relationship superimposed on the observed data.  The~most apparent behavior of the model grid is a progressive downward shift with aging, i.e.,~an intensive decline in $\OIII \lambda5007/\mathrm{H}\beta$ relative to a weak change in $\NII\lambda6583/\mathrm{H}\alpha$.  It is because the \OIII~emission lines are more preferentially excited at a higher degree of ionization, in~contrast to the $\NII\lambda6583$ emission lines associated with low-ionization environments (\citet{OsterbrockAstrophysics, Kewley2019}); with the aging of stellar populations and resultant softening of stellar radiations, \HII~regions have lower ionization states and more delicate \OIII~excitation.  
A~direct aftermath of the change in the spectral features is a decrease in the coverage rate of the model to the data in the diagram.  The~model grids with 0--1\,Myr cover $\sim$80\% of the data points, but~the coverage rate rapidly reduces to $<$30\% at $\geq$2\,Myr and 0\% (no coverage) at $\geq$$6.5$\,Myr.  The~total separation of the model grid from the observed data suggests the upper age limit $\sim$$6.5$\,Myr for the \HII~regions in NGC\,2403 under this framework.  
{We further cross-checked the age constraint using the equivalent width of {$\Hb$} ({$EW_{\Hb}$}) as predicted by the Starburst99 models for instantaneous bursts (\citet{Leitherer1999}), based on the 1999 dataset from the official website\endnote{\url{https://massivestars.stsci.edu/starburst99}, {accessed on 8 August 2025.}
}. For~each metallicity, we compared the model {$EW_{\Hb}$}–age relations with the observed maximum and minimum {$EW_{\Hb}$} values in Figure~\ref{fig:ew_age}. The~smallest {$EW_{\Hb}$} was found in the central high-metallicity regions, corresponding to an age of $ \sim$$7$\,Myr (for $Z=0.02$) in the models. Given that the observed continuum contains contributions from the underlying stellar populations of the galactic disk in addition to the \HII regions, the~measured {$EW_{\Hb}$} is likely underestimated, thus biasing the age estimate toward older values. Taking this effect into account, the~upper limit of $\sim$$6.5$\,Myr inferred from the BPT coverage remains a reasonable estimate.}

{Combining Figures~\ref{fig:BPT_dig},
~\ref{fig:index_grid_0Myr}, \ref{fig:BPT_age} and \ref{fig:index_grid_1.0Myr}--\ref{fig:index_grid_9.5Myr}}, we find that a few of the data points from the central area of the galaxy are missed by the 0\,Myr model but covered by the 2--6\,Myr grids, which indicates an older stage for these \HII~regions than others in the sample.  It is interesting that at the 3--6 Myr evolution phase, the~model grids emerge with peculiar fluctuation at the high-metallicity part, which is ascribed to an influence of Wolf--Rayet (W-R) stars on the \OIII~emission lines (\citet{Levesque2010}).

Variations in the model coverage to the observed data in the index--index diagrams with age are shown in Figure~\ref{fig:CR}, with~an evolutionary trend of the average coverage among all of the diagrams superimposed.  We can see that in general, for almost all of the diagrams, the~model coverage stands at the highest levels at 0--1\,Myr but~drops sharply at 2--6\,Myr and~is lowest after 6\,Myr.  The~highest coverage rates differ between the various diagrams by~$\sim$50\% at most.  All of the $R23$-relative diagrams possess the coverage rate $\leq$$60\%$.  It is suspected that the model grids are likely to have some problem in reproducing the $\OII$ emission line or/and sampling the parameter spaces, which is discussed in Section~\ref{sec:Disc:Oii}.  By~contrast, the~$N2O2$-$O3N2$ and $N2O2$-$O3O2$ curves maintain high coverage rates for a long period ($\geq$$4$\,Myr timescale), because~the ionization parameter and metallicity are more effectively decoupled in the two diagrams, leading to larger areas of the model grids with higher possibilities for covering the data points.  After~6\,Myr, the~coverage rate for most of the diagrams decreases to <10\%, except~for $N2O2$-$N2$ and \NII-\OII, which are able to keep $\sim$$30\%$ coverage rate after 6\,Myr.  This is because the \OII~and \NII~emission lines are low-ionization emission lines not as sensitive to age as others, which is depicted in Figure~\ref{fig:index_age}\endnote{{In Figure~\ref{fig:index_age}, we note that some of the high-metallicity model points (in red and yellow) exhibit abrupt jumps at older ages. We suspect that these anomalies are artifacts arising from the model calculations. However, since this cannot be confirmed and they do not significantly impact our analysis, we chose to retain them.}}.

\begin{figure}[H]
	\centering
	\includegraphics[width=\linewidth]{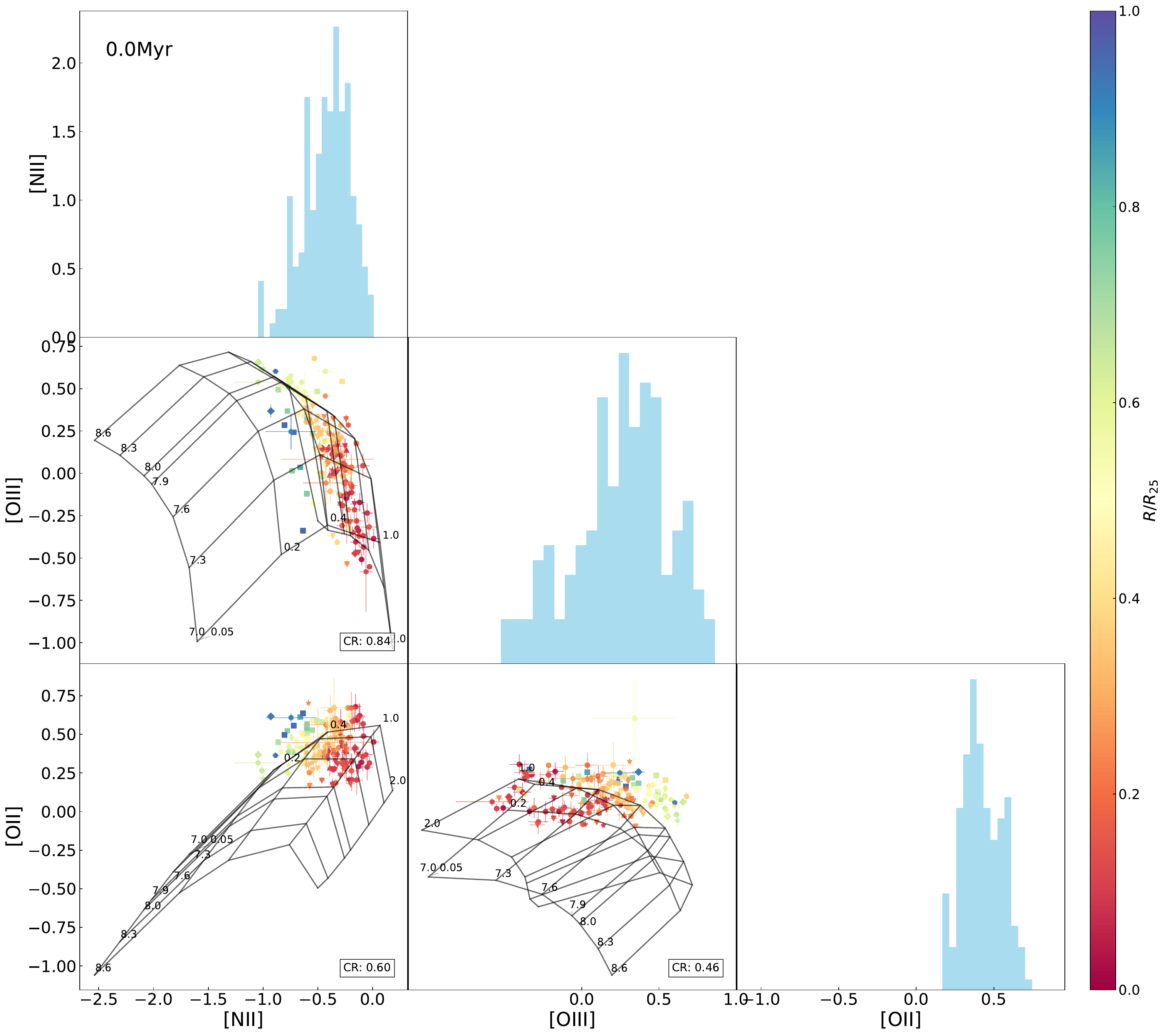}
	\caption{Interrelations between every two emission-line ratio (normalized by \Hb) among \OII, \NII,~and \OIII~for the \HII~regions in NGC\,2403, with~the \citet{Levesque2010} model grid (constant age of 0\,Myr, variations in metallicity Z (in units of $Z_{\bigodot}$) and the ionization parameter q) superimposed.  {The data points are marked using the same symbols as in Figure~\ref{table:slit}} and color-coded the same as in Figure~\ref{fig:BPT_dig}.  The~{CR (same definition as in Figure~\ref{fig:index_grid_0Myr})} is presented in the lower-right corner in each of the panels. Histograms of the indices are also~shown.}
	\label{fig:emission_line_grid_0Myr}
\end{figure}

By modeling $N2O2$, $N2$, $O3N2$, $R23$, $O3O2$, \OII, \OIII, and \NII~at a variety of values for metallicity, the~ionization parameter, and~stellar population age addressed in \citet{Levesque2010}, Figure~\ref{fig:index_age} reproduces the age evolution of the spectral indices and the emission line {ratios}, at~each constant metallicity and ionization parameter, respectively, as~an underlying factor interpreting the above features in the diagrams (\mbox{{{Figures} 
 \ref{fig:index_grid_0Myr}}, \ref{fig:BPT_age}, \ref{fig:emission_line_grid_0Myr} and~\ref{fig:index_grid_1.0Myr}--\ref{fig:emission_line_grid_9.5Myr}}).  We can see from the evolutionary tracks that the indices $N2O2$ and $N2$ are the least sensitive to age; $N2O2$ even appears invariant with respect to the ionization parameter in the whole age range of 0--10\,Myr, suggesting negligible degeneracy (if any) of metallicity,~ionization parameter, and~stellar population age in these indices, but~$N2$ jointly depends on both metallicity and ionization parameter for \HII~regions during the first 6\,Myr.

The $O3N2$ tracks gradually decline with age at lower metallicity than the solar value ($Z \leq 0.02$) but~fluctuate conspicuously in the 3–6\,Myr range at higher metallicities \mbox{($Z = 0.04$)}; $O3N2$ is also sensitive to both metallicity and ionization parameter, similar to $N2$ but spanning the whole 0--10\,Myr range.  $O3O2$ is similar to $O3N2$ in terms of age tracks but~more dependent on the ionization parameter than metallicity, and~this is the reason for this index being commonly adopted as an ionization proxy.  The~$R23$ index exhibits a weak declining trend with age and~a decreasing sensitivity to the ionization parameter as metallicity increases.  The~fluctuation in the 3–6\,Myr range is a noticeable feature in all of the tracks for $O3N2$, $O3O2$, and $R23$ at the metallicity $Z = 0.04$, coincident with the \OIII~tracks at the same age and metallicity.  As~explained above, this feature is caused by the influence of W-R stars on the \OIII~emission line, which results in the peculiar behaviors of the 3–6\,Myr model grids in all of the \OIII~relative diagrams and has potential impact on observed data loci in this special age interval.  The~relation with age is strong for \OIII~but weak for \OII~and \NII~emission lines, which suggests strong age dependence of $O3N2$ and $O3O2$ but~insensitivity to $N2O2$ and $N2$.

\section{Discussion} \label{sec:Discussion}
\unskip

\subsection{Effects of~Apertures}\label{sec:Disc:aper}

Differences in position and size of apertures for spectral extraction have much potential to introduce extra uncertainties in the measurements.  The~specific effects of apertures on different emission lines and different spectral indices are suspected to be different.  For~instance, higher-ionization emission lines such as \OIII~tend to be more compact and concentrated in central parts of \HII~regions, whereas lower-ionization emission lines such as \OII~and \NII~often extend over a wider scale (\citet{Kewley2019}); for the behaviors of spectral indices, spatially resolved observations of large \HII~regions have revealed that $O3O2$ peaks around the nebular centers, while $N2O2$ is relatively invariable across different locations within the \HII~regions (\citet{Mao2018}).
 
 As a consequence, if~apertures are more centrally placed, they are more likely to capture stronger \OIII~emission lines, leading to higher values for $O3O2$.~By~contrast, larger apertures enclosing more outer areas of \HII~regions are more likely to extract stronger \OII~and \NII~emission lines and hence obtain a decrease in $O3O2$ or $O3N2$.

\subsection{Potential Limitation in the Model on the {\OII} 
~Emission~Line}\label{sec:Disc:Oii}

In the above results, the~model fails to adequately fit the observed data in the $N2O2$-$N2$ diagram as well as $R23$-relative diagrams, particularly for the \HII~regions located in outskirts of the galaxy with relatively low metallicities.  In~order to find an interpretation, we further inspected line--line diagrams, as shown in Figure~\ref{fig:emission_line_grid_0Myr}. We can see that the model grid covers almost all of the data points in the $\OIII$-$\NII$ diagram but only one half in the $\NII$–$\OII$ and $\OIII$–$\OII$ diagrams.  Variations in age are sufficient to account for the low coverage in the $\NII$–$\OII$ and $\OIII$–$\OII$ diagrams (see Figure~\ref{fig:CR}).

The parameter space of metallicity and the ionization parameter in the model, in~principle, is supposed to cover the observed data.  It is unclear whether the failure of the coverage in the $\OIII$–$\OII$ diagram is caused by inadequate sampling resolution or other inherent limitations in the model.  From~the $\NII$–$\OII$ diagram, we can affirm that the \citet{Levesque2010} model grids are unable to reproduce the observed lines at low metallicity end typically in the galactic outskirts.  Specifically, for~a given $\NII$, the~model fails to reproduce a sufficient range of $\OII$ intensities to match the observed data.  This limitation also provides an interpretation of the lack of model coverage in the $N2$–$N2O2$ diagrams as presented above.  Indeed, generation of certain emission lines in photoionization modeling is an intrinsically complicated mechanism.  In~addition to metallicity, the~ionization parameter, and~gas density, many other parameters, such as the shape and the intensity of ionizing radiation field, as~well as the gas geometry and physics, play critical roles in predicting emission~lines.

\subsection{Impact of Rotation of Stars or Binary~Populations}\label{sec:Disc:RotBinary}

Spectral types of ionizing sources are a dominant factor for determining emission lines in photoionization processes and dependent on several assumptions in stellar population synthesis such as initial mass function, star formation history, stellar evolutionary tracks, stellar rotation,~binary-population interactions, etc.  It is mentioned in \mbox{\citet{Levesque2010}} that the model is likely to require harder spectra to create sufficient ionizing fields and yield enough \SII~to successfully reproduce observed data.  In~this case, rotation of stars is suggested to achieve this.  The~stellar rotation is expected to create harder ultraviolet spectra for a longer period of time \mbox{(\citet{Levesque2012ApJ...751...67L,Dorn-Wallenstein2020ApJ...896..164D})}.  Alternatively, binary stars are another possible requisite for maintaining hardness of ultraviolet radiation despite aging stellar populations {(\citet{Stanway2016MNRAS.456..485S})}.  Binary stellar population models make it possible to continuously reproduce observed data after 10\,Myr in the BPT diagram and~those at low metallicity or high redshifts in the \OIII-\SII~diagram (\citet{Zhang2015MNRAS.447L..21Z, Xiao2018MNRAS.477..904X}).

\section{Summary} \label{sec:Summary}

In this paper, we presented an investigation of spectral emission-line characteristics of \HII~regions in the nearby spiral galaxy NGC\,2403.  Spectroscopic observations were taken with the 2.16\,m telescope at the National Astronomical Observatories of China, with~supplementary data collected from the literature.  As~a consequence, a~total of 188 \HII~regions in NGC\,2403 were compiled in the final sample.  A~combined model of photoionization and stellar population synthesis was employed for comparing and interpreting the observed~data.

Through the BPT diagnosis, {nearly all of the samples} were confirmed to originate from stellar irradiation and hence classified as star-forming regions.  Radial profiles of metallicity for NGC\,2403 were obtained by applying the four widely used strong-line diagnostics, $R23$, $O3N2$, $N2$, and $N2O2$, respectively, and~all of them exhibited negative gradients, reflecting an inside-out evolution process for this galaxy, but~the slopes and the dispersions of the gradients were different. $N2O2$ resulted in the gradient with the steepest slope and the least dispersion, while the $N2$ {and $R23$} indicators yielded the flattest gradients, and~$R23$ exhibited the largest~dispersion.

Via a series of diagrammatic inspections on the four spectral indices $R23$, $O3N2$, $N2$, and $N2O2$, as well as individual emission lines within them, we demonstrated strong sensitivities of $O3N2$ and $N2$ to the ionization parameter; $R23$ was comparatively less sensitive to the ionization parameter but had a double-value effect which was especially influential for the \HII~regions in NGC\,2403; $N2O2$ appeared to be the most robust diagnostic, in~monotonic correlation with metallicity and with negligible sensitivity to the ionization parameter. By~inspecting the individual emission lines, we disclosed that the \OIII~emission line was strongly related with the age of ionizing stellar populations, in~contrast to the \OII~or \NII~emission line in relatively weak relation with stellar population age. As~a consequence, the~indices $R23$ and $O3N2$ involving the \OIII~emission line were dependent on stellar population age, while $N2$ and $N2O2$ were less affected by stellar population age, but~the influence of the ionization parameter on $N2$ became more serious for young stellar populations ($\lesssim$6~Myr). Thus, $N2O2$ is suggested to more reliably indicate metallicity than the other three diagnostics. Furthermore, within~the adopted photoionization model framework, the~grid corresponding to an age of 0–1 Myr provided the best overall match to the observational data, whereas models older than 6.5 Myr showed poor~agreement.

\vspace{6pt} 


\authorcontributions{{Conceptualization, Q.-M.W. and Y.-W.M.; Methodology, Q.-M.W. and \mbox{Y.-W.M.;} Software, Q.-M.W.; Validation, Q.-M.W. and Y.-W.M.; Formal analysis, Q.-M.W.; Investigation, Q.-M.W. and Y.-W.M.; Resources, Y.-W.M. and L.L.; Data curation, Q.-M.W., Y.-W.M. and S.-T.W.; writing---original draft preparation, Q.-M.W. and Y.-W.M.; writing---review and editing, Q.-M.W., Y.-W.M., L.L., H.Z. and S.-T.W.; Visualization, Q.-M.W.; Supervision, Y.-W.M.; Project administration, Y.-W.M.; Funding acquisition, Y.-W.M., L.L. and H.Z. All authors have read and agreed to the published version of the manuscript.}}

\funding{{This work is supported by the National Key R\&D Program of China (No. 2022YFA1602902), the~National Natural Science Foundation of China (NSFC, Nos. U2031106, U2031201, 12120101003, and~12373010), China Manned Space Project (Nos. CMS-CSST-2025-A06 and CMS-CSST-2021-B04), and~the Youth Innovation Promotion Association CAS (id. 2022260).   }}

\acknowledgments{ 
 We acknowledge the support of the staff of the Xinglong 2.16 m telescope. This work was partially supported by the Open Project Program of the Key Laboratory of Optical Astronomy, National Astronomical Observatories, Chinese Academy of Sciences. This research has made use of the NASA/IPAC Extragalactic Database (NED) which is funded by the National Aeronautics and Space Administration and operated by the California Institute of Technology; this research has also made use of the SAO/NASA’s Astrophysics Data System Bibliographic Services.
 }
 
 \conflictsofinterest{{The authors declare no conflicts of interest.}}

\appendixtitles{yes} 
\label{sec:App_index_age}

\appendixstart{
\appendix


\section[\appendixname~\thesection]{Index--Index Diagrams with Age Evolution}\label{sec:App_index_age}

In this appendix, we present the index--index diagrams and the line--line diagrams, with~the model grids at different ages superimposed, corresponding to the analysis in Section~\ref{sec:diag_age}.  The~index--index diagrams are shown in Figures~\ref{fig:index_grid_1.0Myr}--\ref{fig:index_grid_9.5Myr}, and~the line--line diagrams are shown in Figures~\ref{fig:emission_line_grid_1.0Myr}--\ref{fig:emission_line_grid_9.5Myr}.

\begin{figure}[H]
  
  \includegraphics[width=0.9\textwidth]{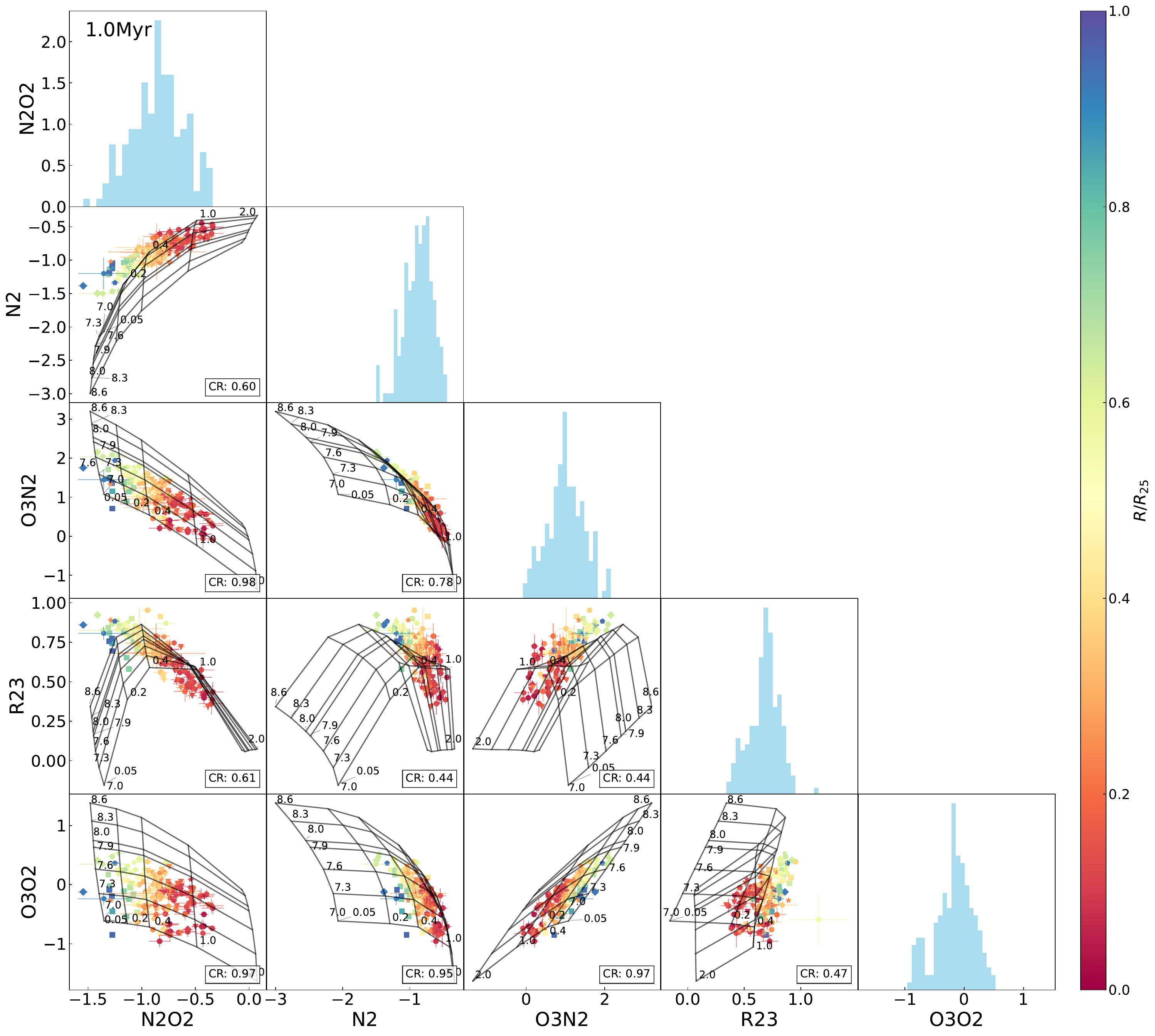}
  \caption{{The} 
 same as Figure~\ref{fig:index_grid_0Myr} but~superimposing the 1.0\,Myr model~grids.}
  \label{fig:index_grid_1.0Myr}
\end{figure}
\unskip

\begin{figure}[H]
  
  \includegraphics[width=0.9\textwidth]{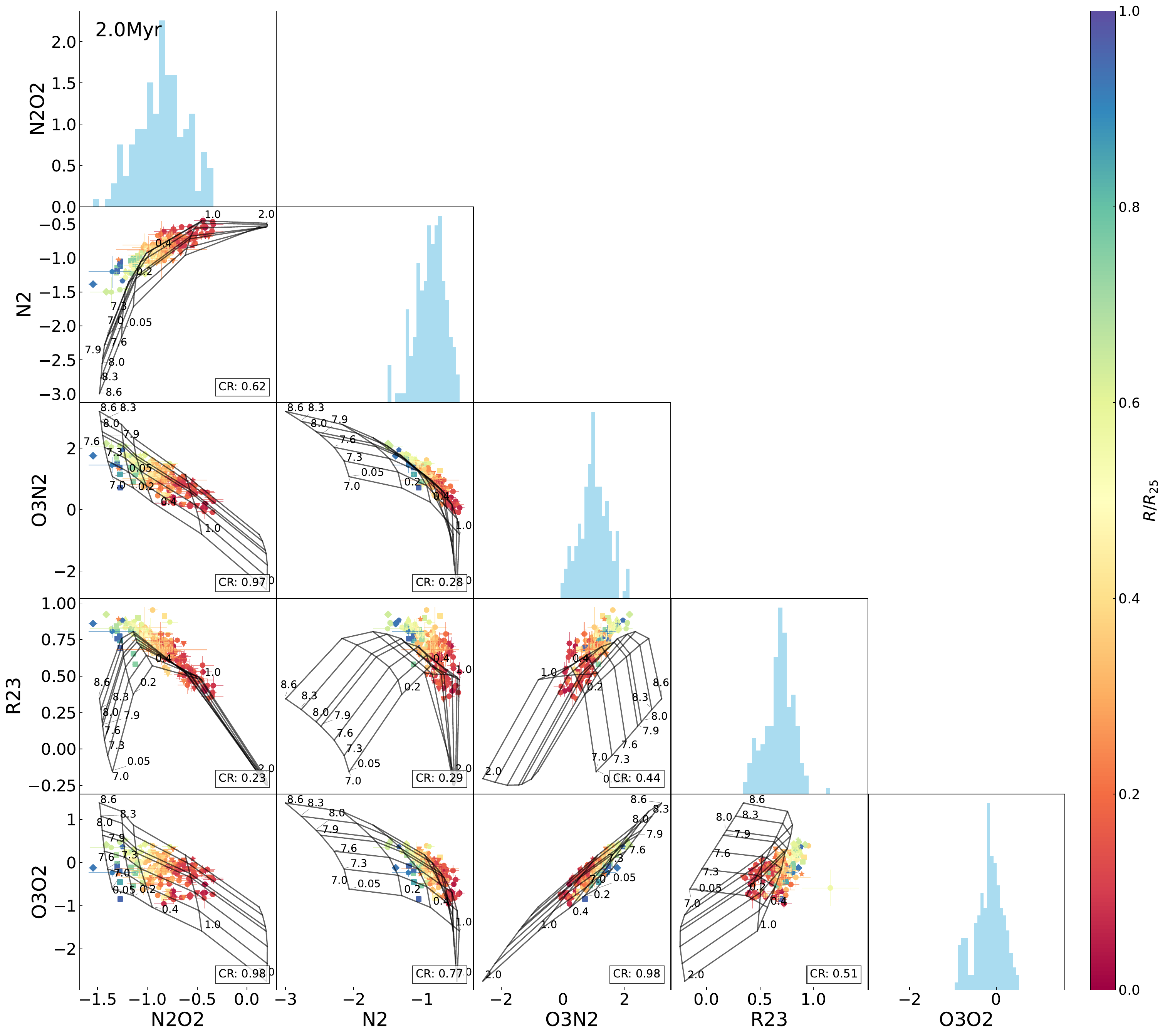}

  \caption{{The} 
 same as Figure~\ref{fig:index_grid_0Myr} but~superimposing the 2.0\,Myr model~grids.}
  \label{fig:index_grid_2.5Myr}
\end{figure}
\unskip

\begin{figure}[H]
  
  \includegraphics[width=0.9\textwidth]{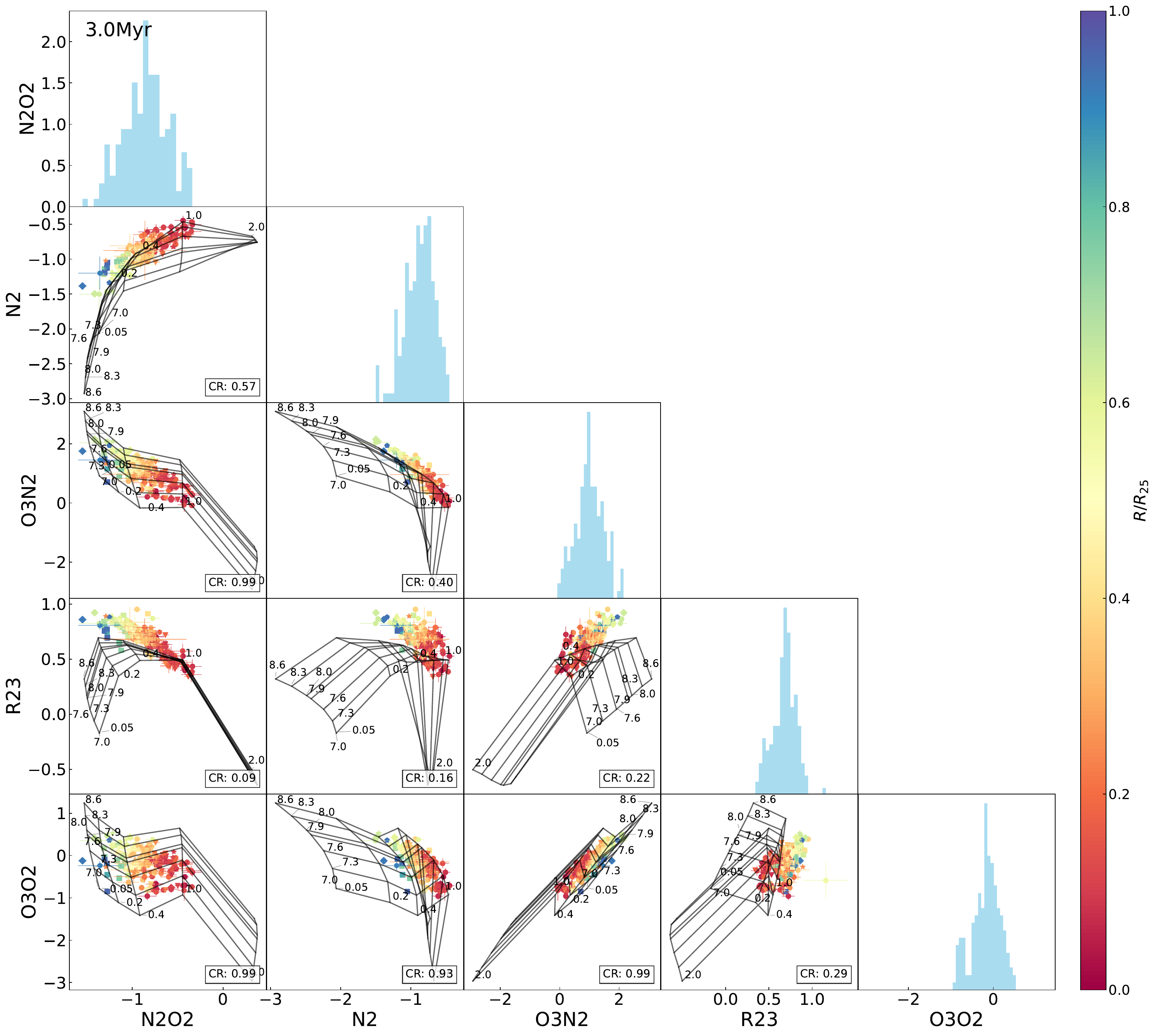}

  \caption{{The} 
 same as Figure~\ref{fig:index_grid_0Myr} but~superimposing the 3.0\,Myr model~grids.}
  \label{fig:index_grid_3.0Myr}
\end{figure}
\unskip

\begin{figure}[H]
  
  \includegraphics[width=0.9\textwidth]{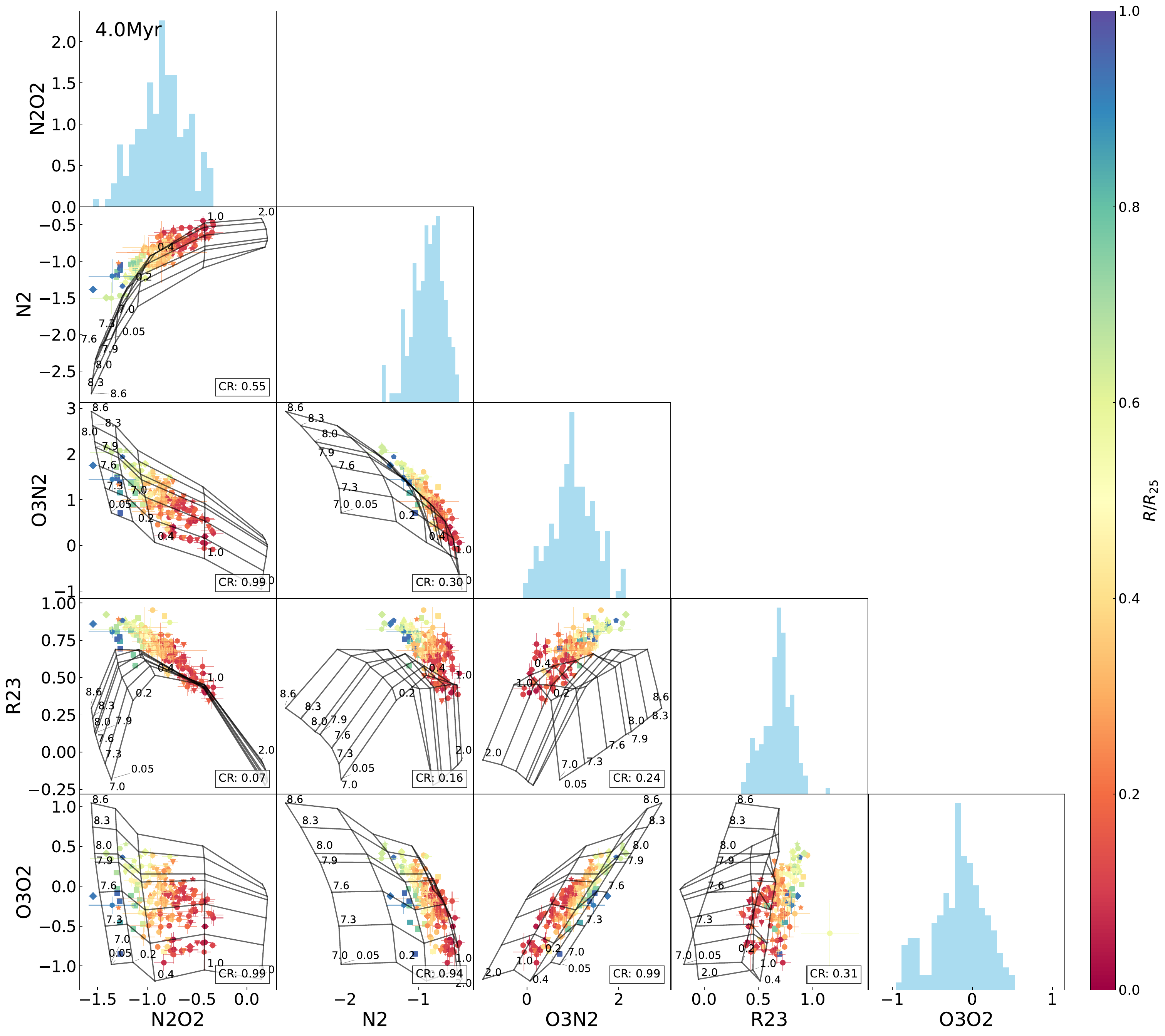}

  \caption{{The} 
 same as Figure~\ref{fig:index_grid_0Myr} but~superimposing the 4.0\,Myr model~grids.}
  \label{fig:index_grid_4.0Myr}
\end{figure}
\unskip

\begin{figure}[H]
  
  \includegraphics[width=0.9\textwidth]{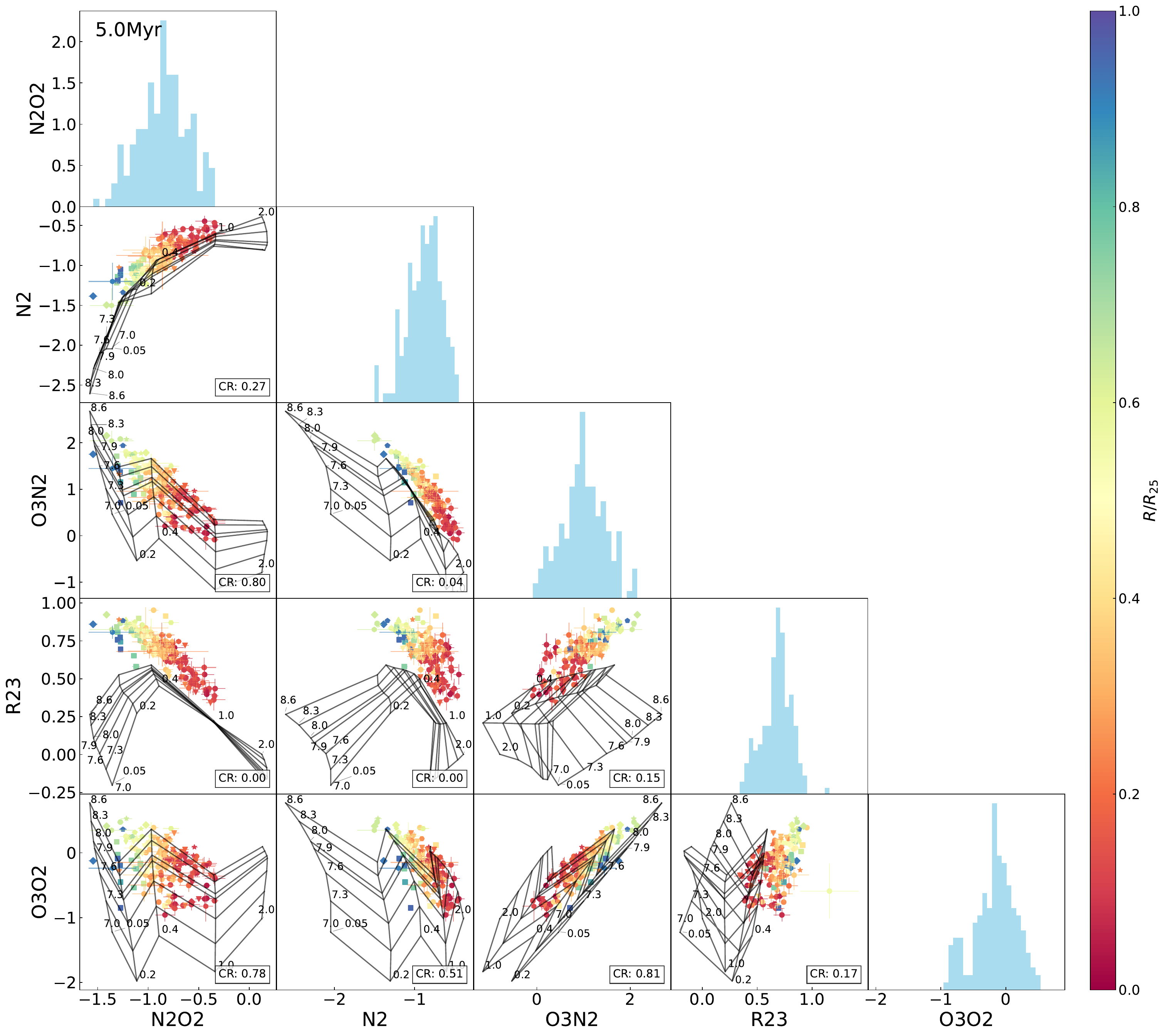}

  \caption{{The} 
 same as Figure~\ref{fig:index_grid_0Myr} but~superimposing the 5.0\,Myr model~grids.}
  \label{fig:index_grid_5.0Myr}
\end{figure}
\unskip

\begin{figure}[H]
  
  \includegraphics[width=0.9\textwidth]{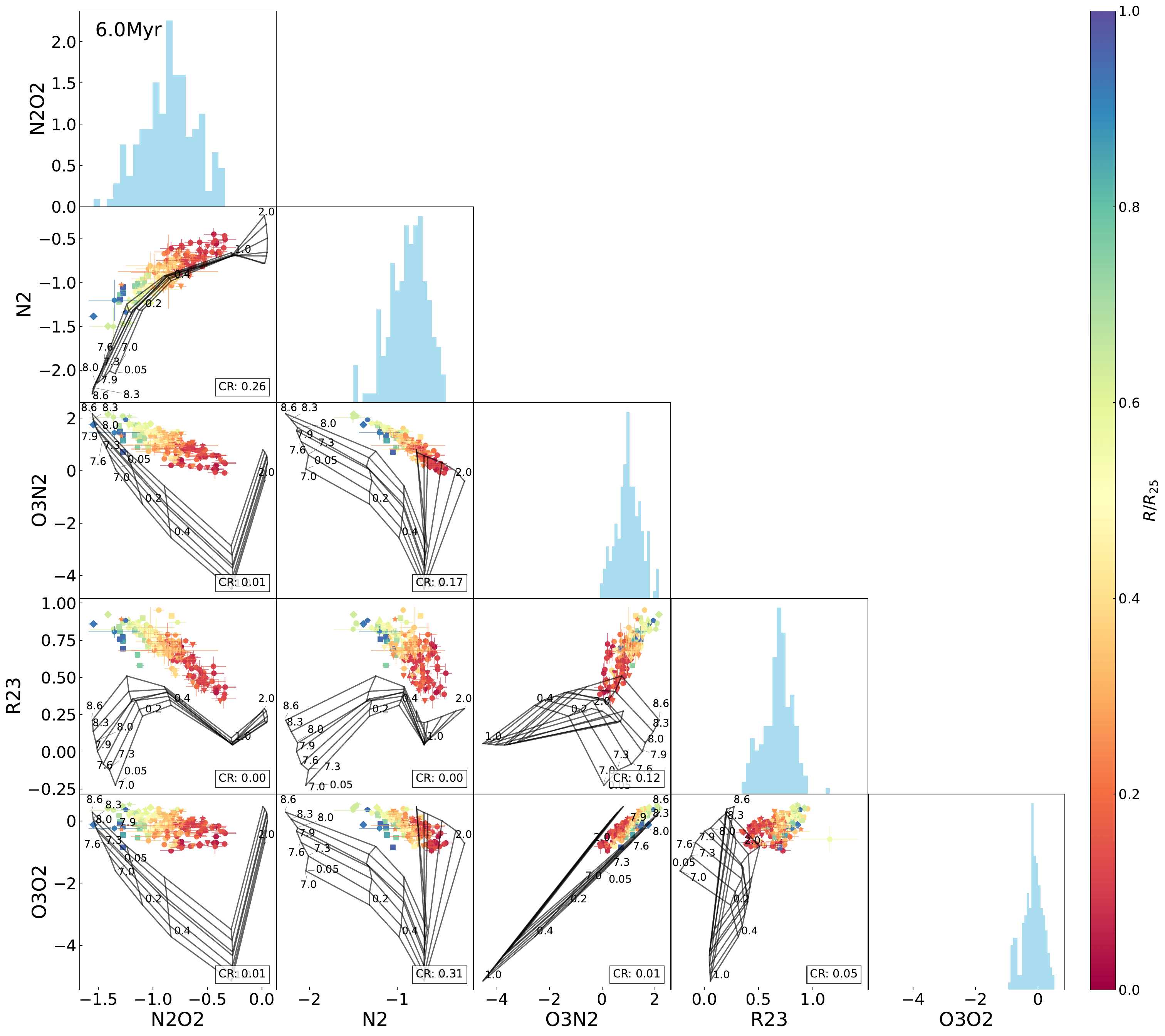}

  \caption{{The} 
 same as Figure~\ref{fig:index_grid_0Myr} but~superimposing the 6.0\,Myr model~grids.}
  \label{fig:index_grid_6.0Myr}
\end{figure}
\unskip

\begin{figure}[H]
  
  \includegraphics[width=0.9\textwidth]{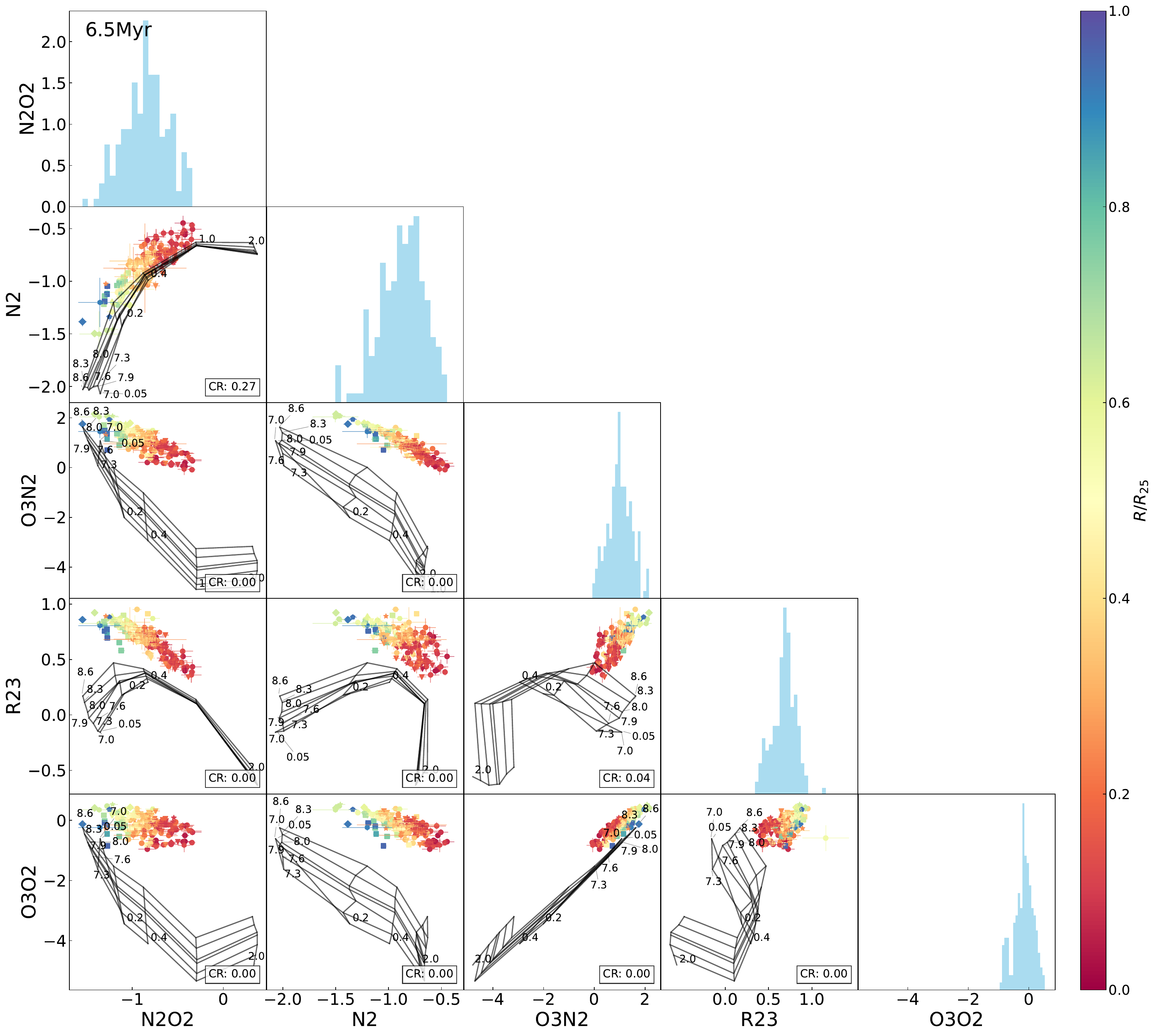}

  \caption{{The} 
 same as Figure~\ref{fig:index_grid_0Myr} but~superimposing the 6.5\,Myr model~grids.}
  \label{fig:index_grid_6.5Myr}
\end{figure}
\unskip

\begin{figure}[H]
  
  \includegraphics[width=0.9\textwidth]{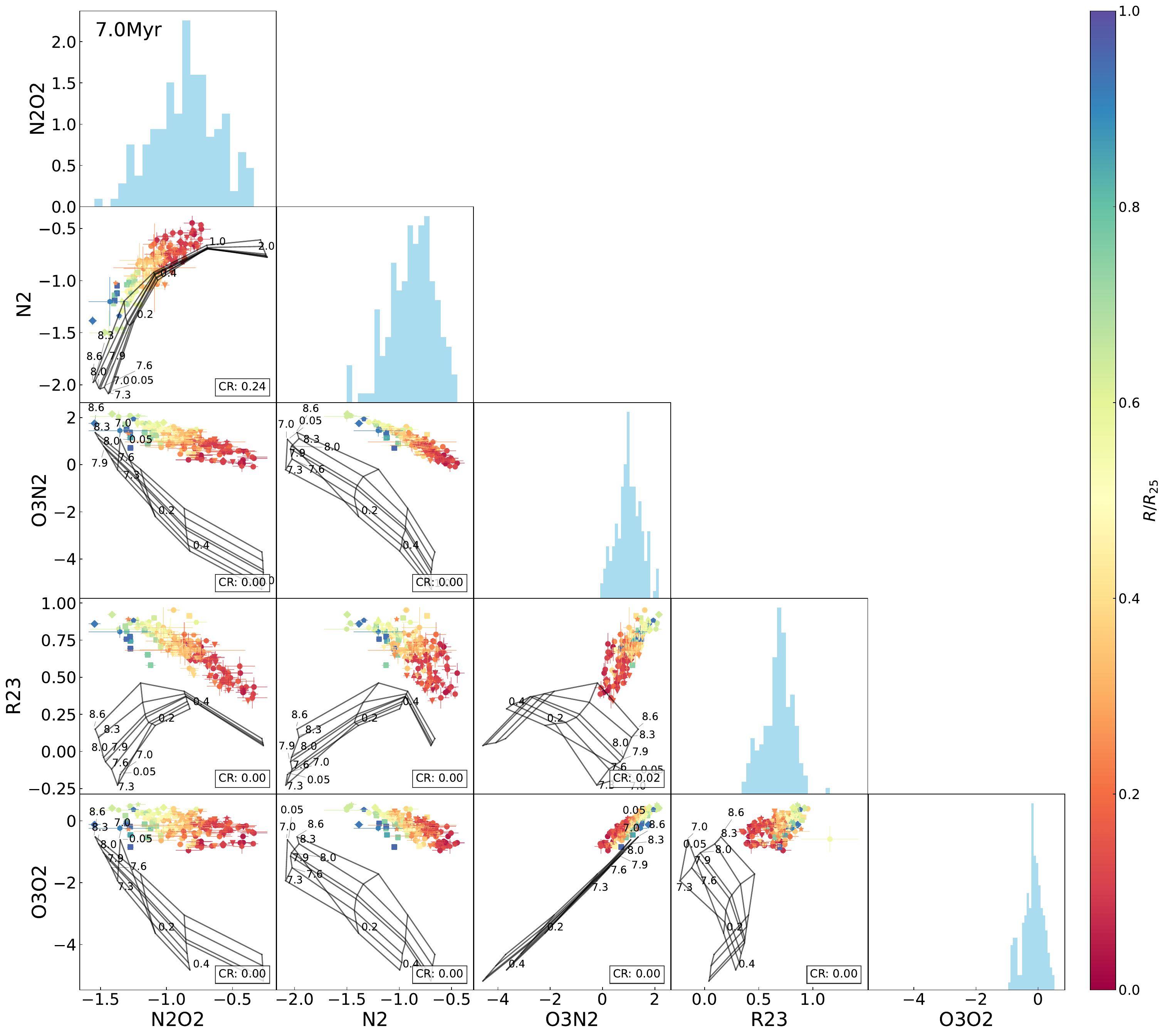}

  \caption{{The} 
 same as Figure~\ref{fig:index_grid_0Myr} but~superimposing the 7.0\,Myr model~grids.}
  \label{fig:index_grid_7.0Myr}
\end{figure}
\unskip

\begin{figure}[H]
  
  \includegraphics[width=0.9\textwidth]{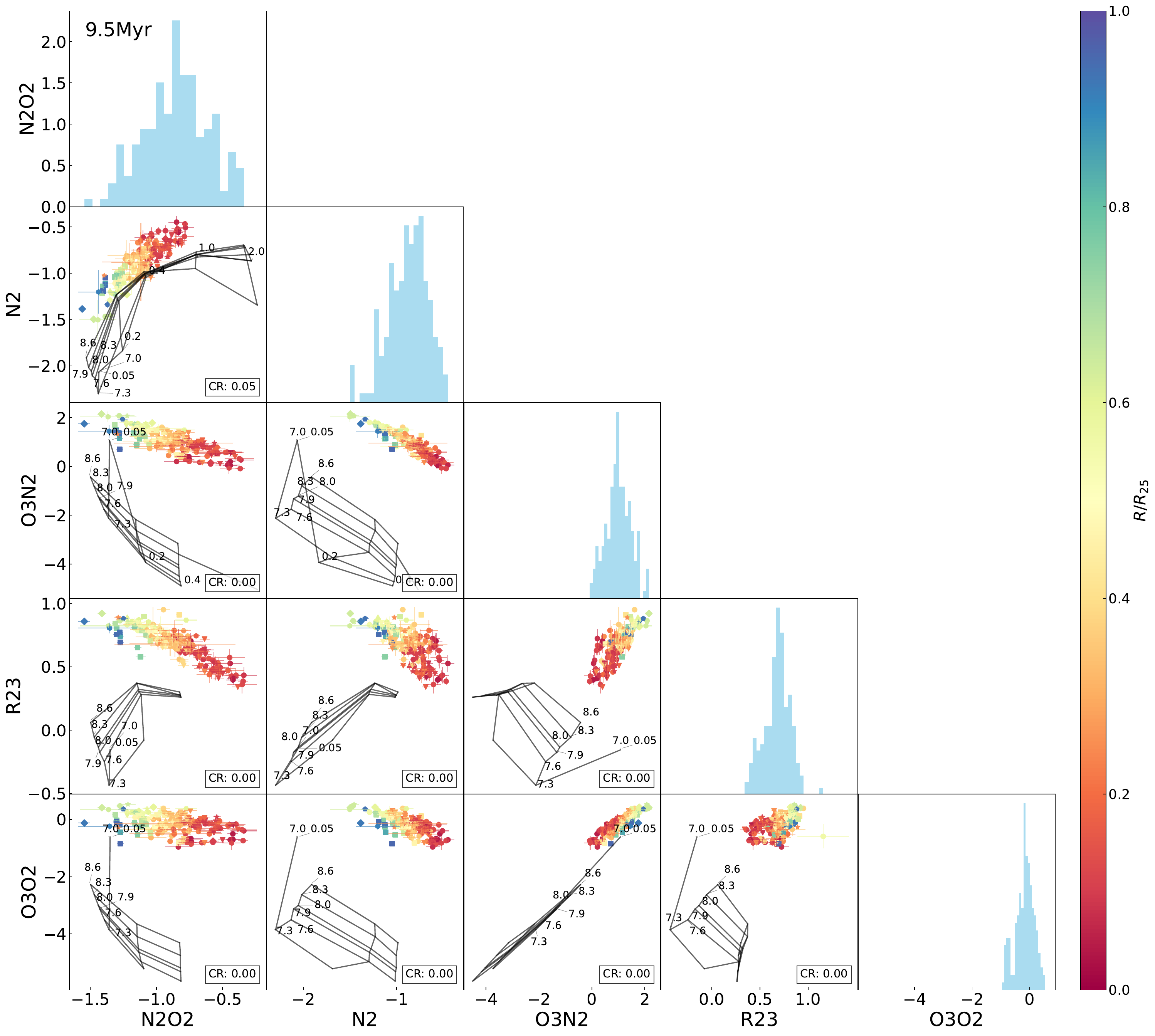}

  \caption{{The} 
 same as Figure~\ref{fig:index_grid_0Myr} but~superimposing the 9.5\,Myr model~grids.}
  \label{fig:index_grid_9.5Myr}
\end{figure}
\unskip

\begin{figure}[H]
  
  \includegraphics[width=0.9\textwidth]{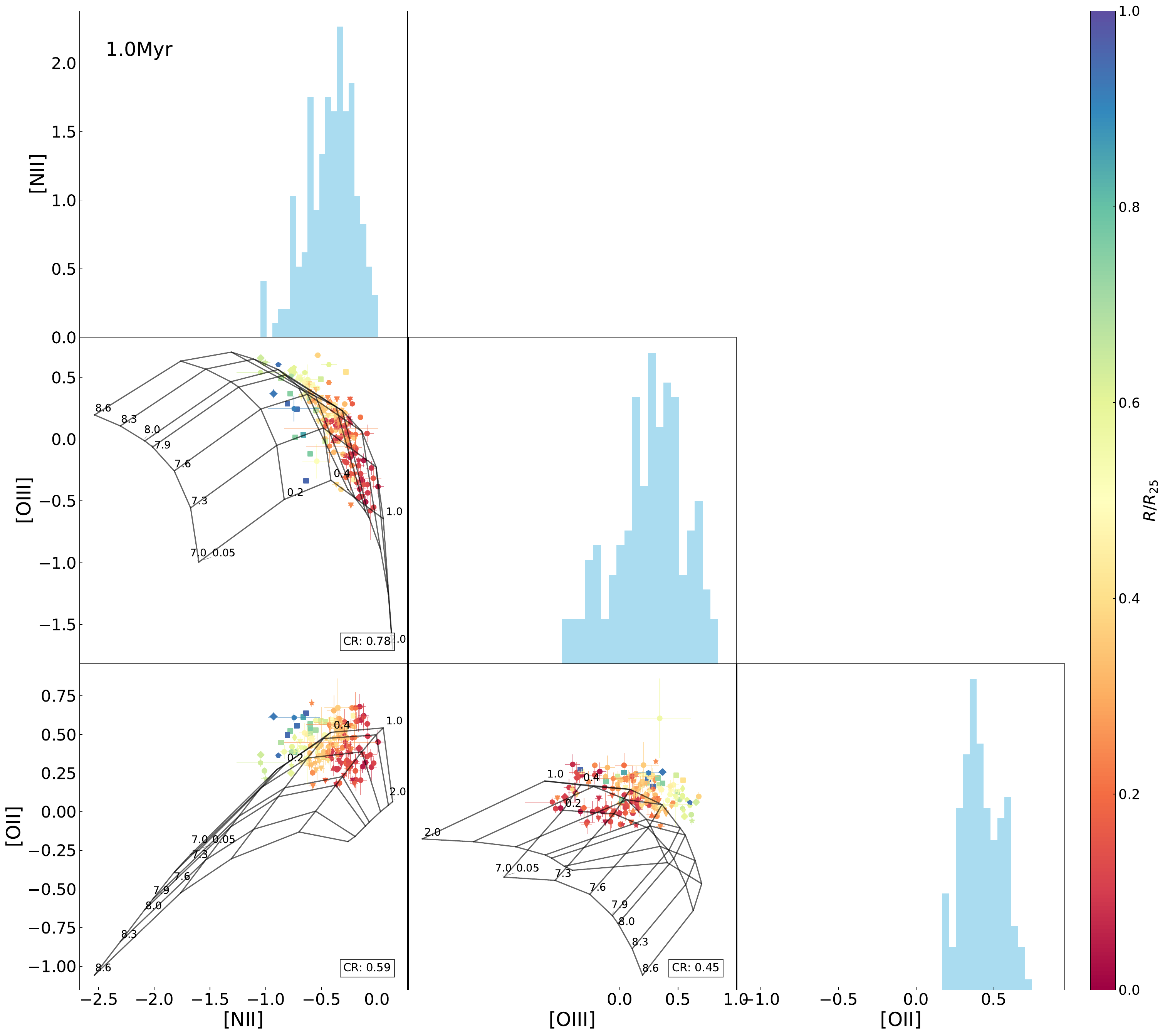}
  \caption{{The} 
 same as Figure~\ref{fig:emission_line_grid_0Myr} but~superimposing the 1.0\,Myr model~grids.}
  \label{fig:emission_line_grid_1.0Myr}
\end{figure}
\unskip

\begin{figure}[H]
  
  \includegraphics[width=0.9\textwidth]{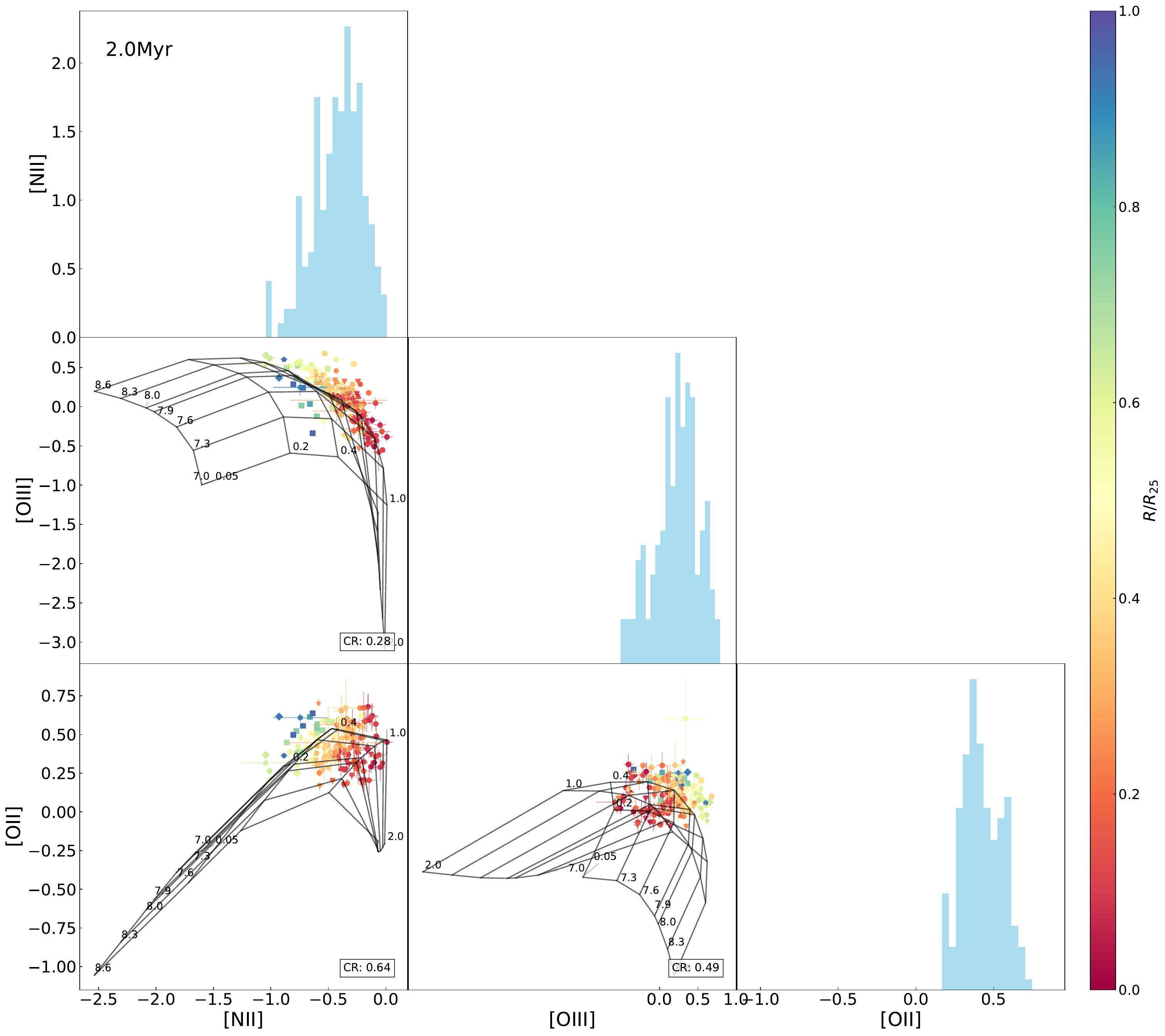}

  \caption{{The} 
 same as Figure~\ref{fig:emission_line_grid_0Myr} but~superimposing the 2.0\,Myr model~grids.}
  \label{fig:emission_line_grid_2.5Myr}
\end{figure}
\unskip

\begin{figure}[H]
  
  \includegraphics[width=0.9\textwidth]{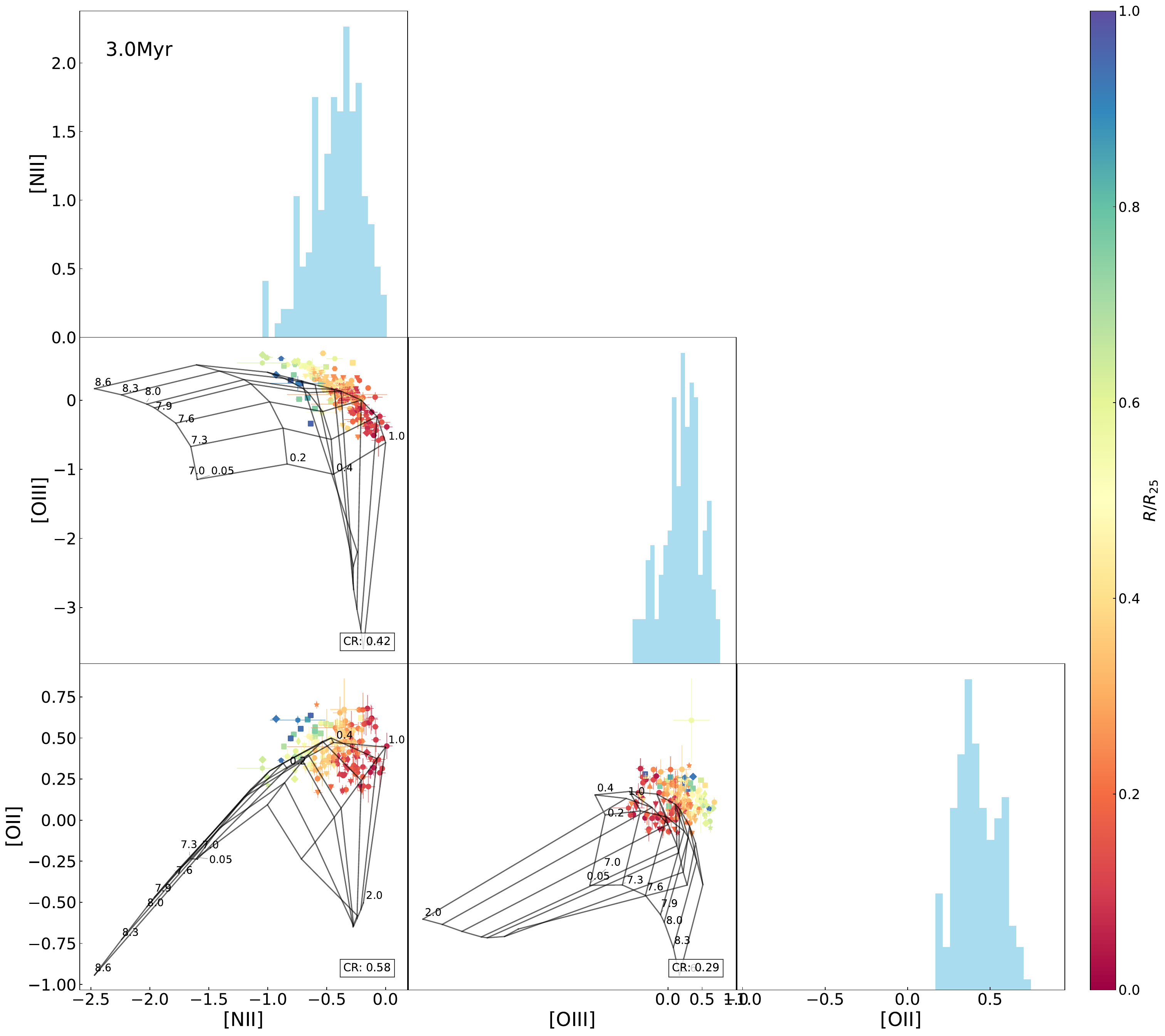}

  \caption{{The} 
 same as Figure~\ref{fig:emission_line_grid_0Myr} but~superimposing the 3.0\,Myr model~grids.}
  \label{fig:emission_line_grid_3.0Myr}
\end{figure}
\unskip

\begin{figure}[H]
  
  \includegraphics[width=0.9\textwidth]{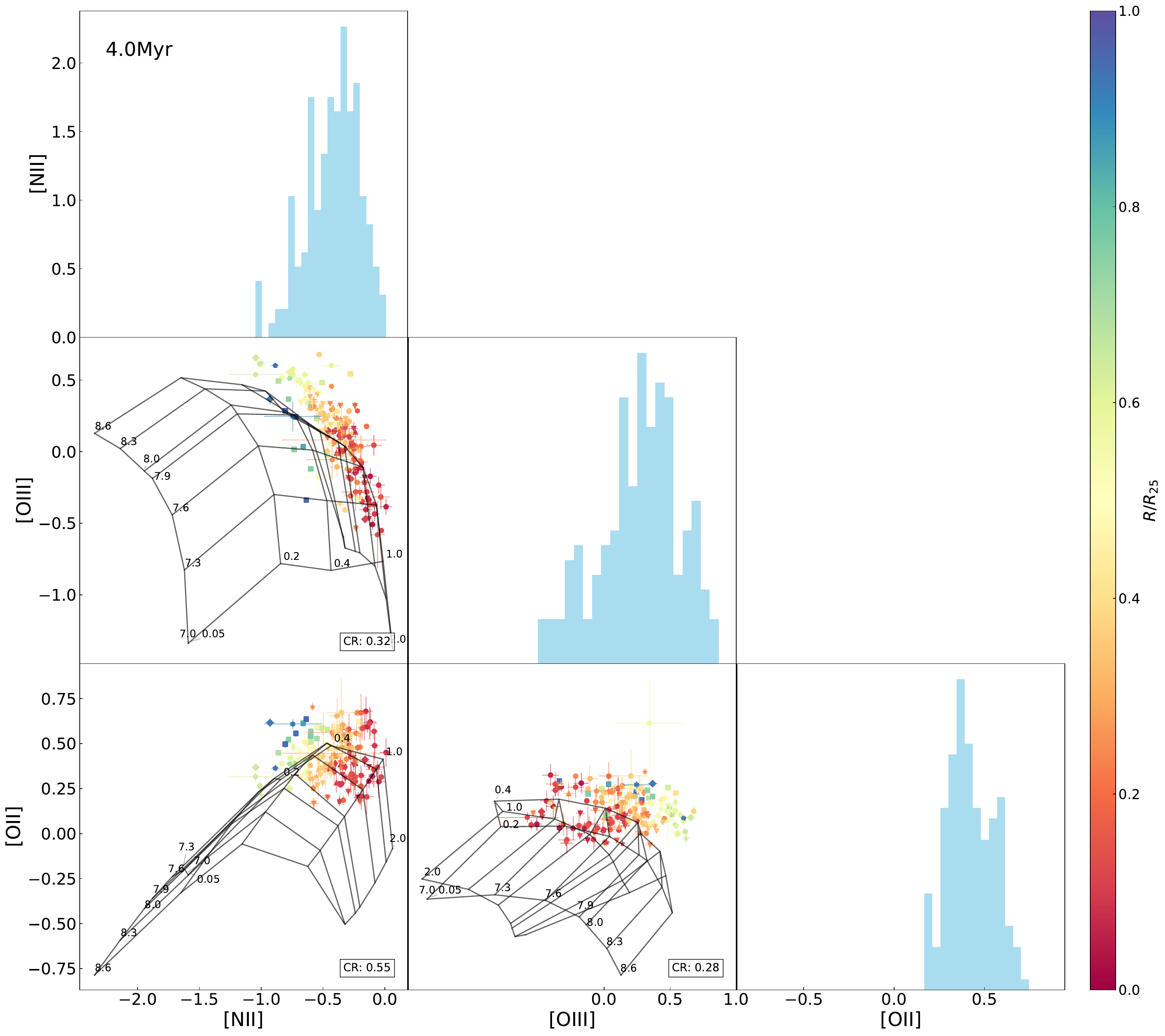}

  \caption{{The} 
 same as Figure~\ref{fig:emission_line_grid_0Myr} but~superimposing the 4.0\,Myr model~grids.}
  \label{fig:emission_line_grid_4.0Myr}
\end{figure}
\unskip

\begin{figure}[H]
  
  \includegraphics[width=0.9\textwidth]{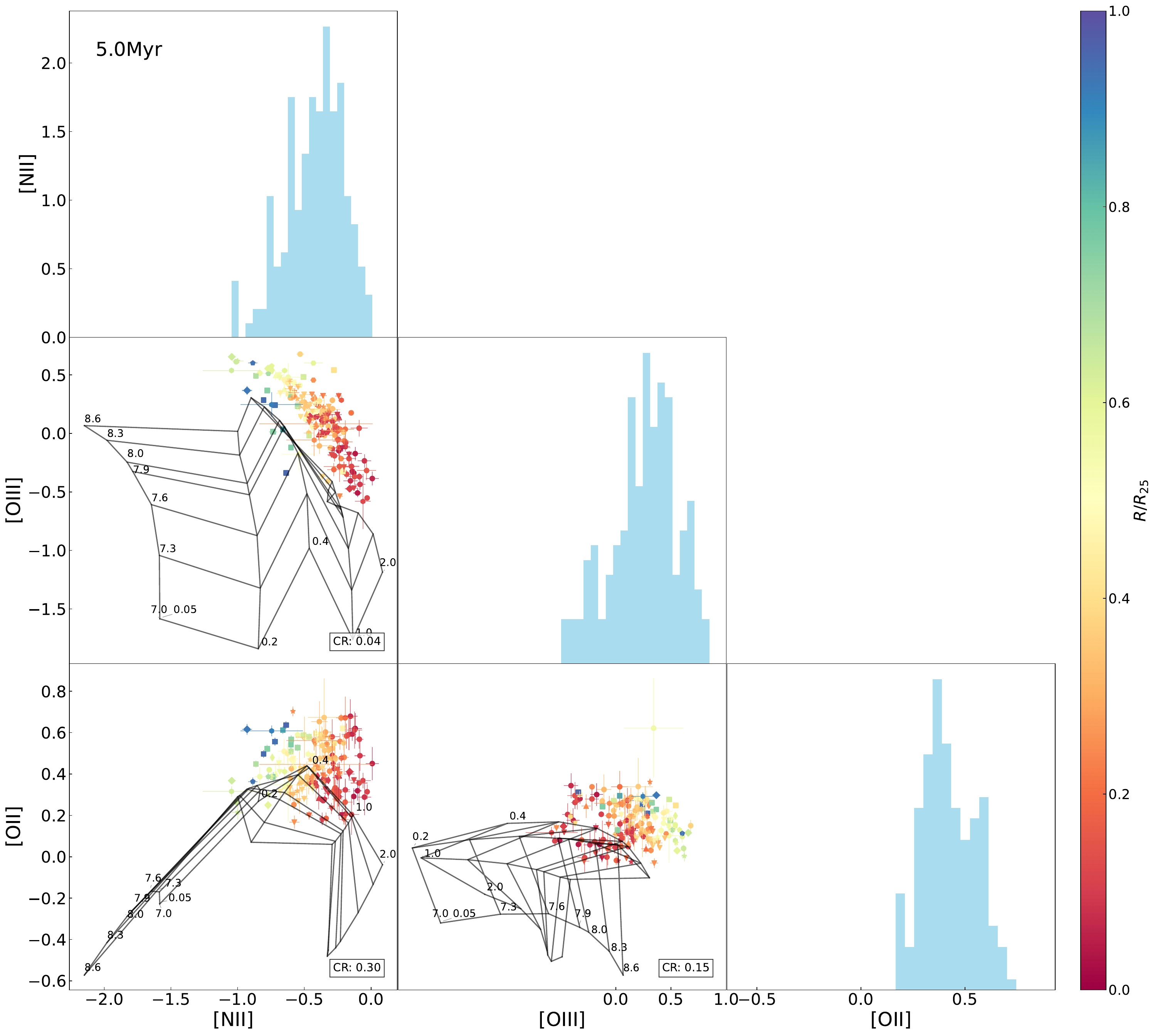}

  \caption{{The} 
 same as Figure~\ref{fig:emission_line_grid_0Myr} but~superimposing the 5.0\,Myr model~grids.}
  \label{fig:emission_line_grid_5.0Myr}
\end{figure}
\unskip

\begin{figure}[H]
  
  \includegraphics[width=0.9\textwidth]{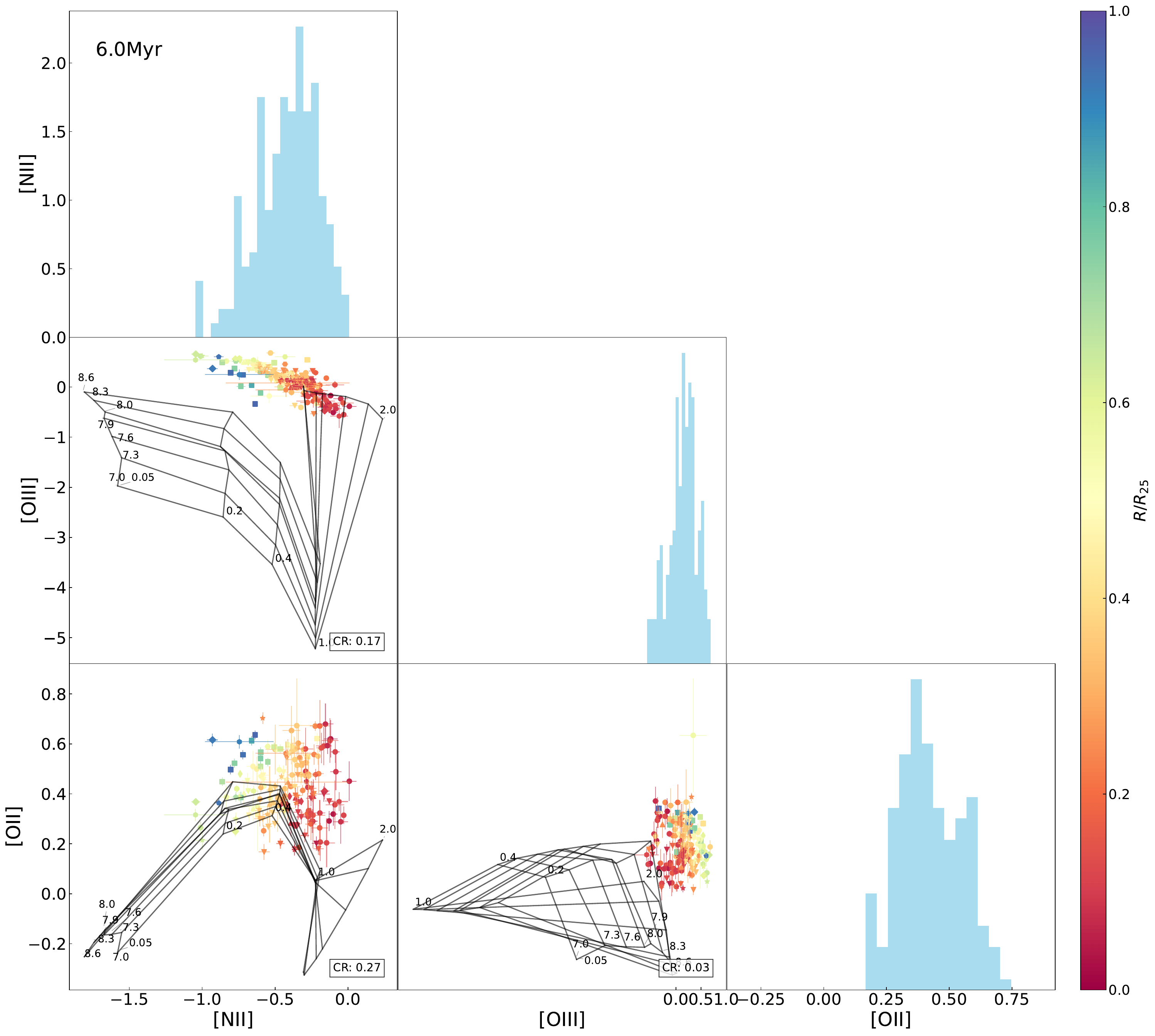}

  \caption{{The} 
 same as Figure~\ref{fig:emission_line_grid_0Myr} but~superimposing the 6.0\,Myr model~grids.}
  \label{fig:emission_line_grid_6.0Myr}
\end{figure}
\unskip

\begin{figure}[H]
  
  \includegraphics[width=0.9\textwidth]{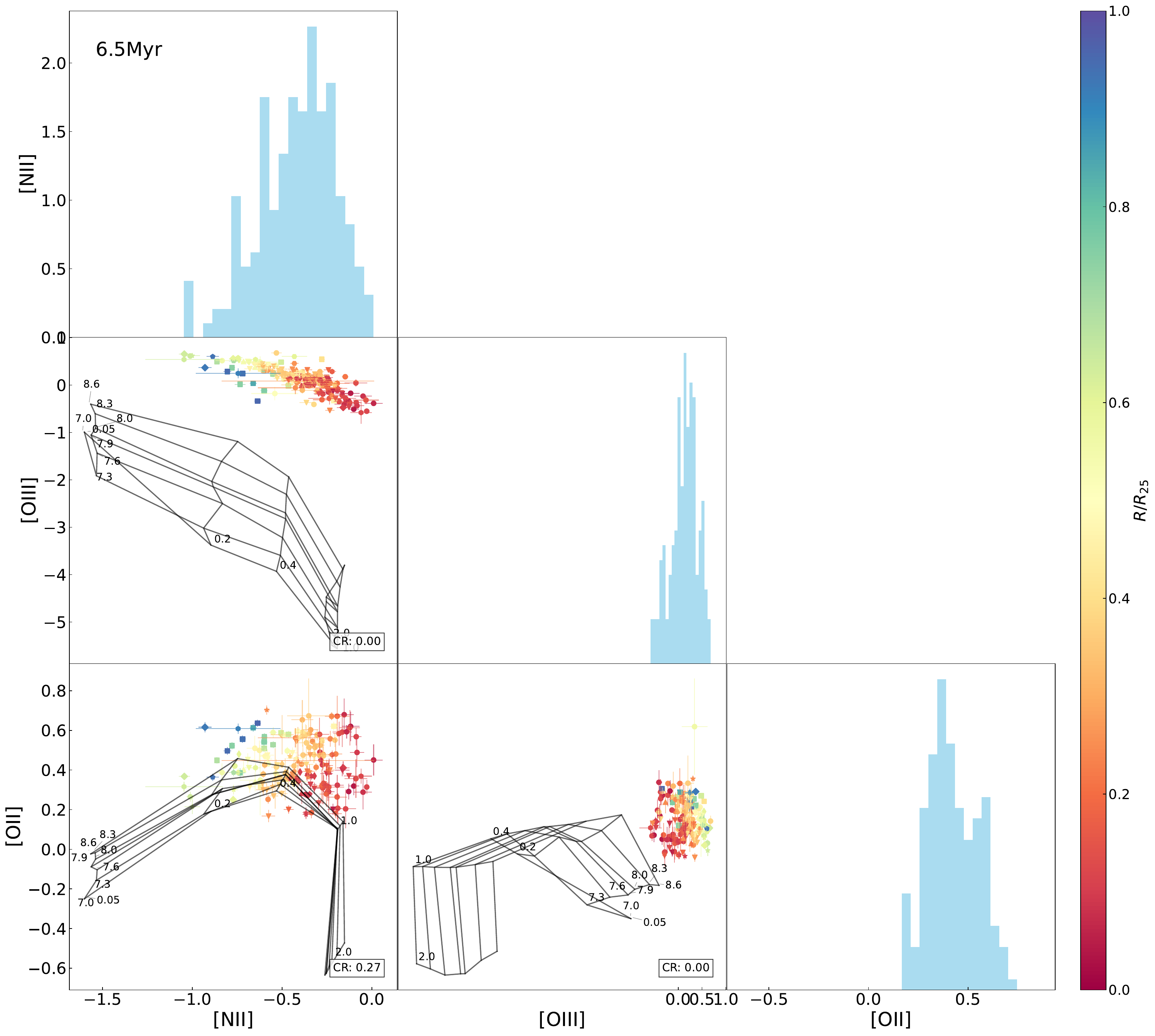}

  \caption{{The} 
 same as Figure~\ref{fig:emission_line_grid_0Myr} but~superimposing the 6.5\,Myr model~grids.}
  \label{fig:emission_line_grid_6.5Myr}
\end{figure}
\unskip

\begin{figure}[H]
  
  \includegraphics[width=0.9\textwidth]{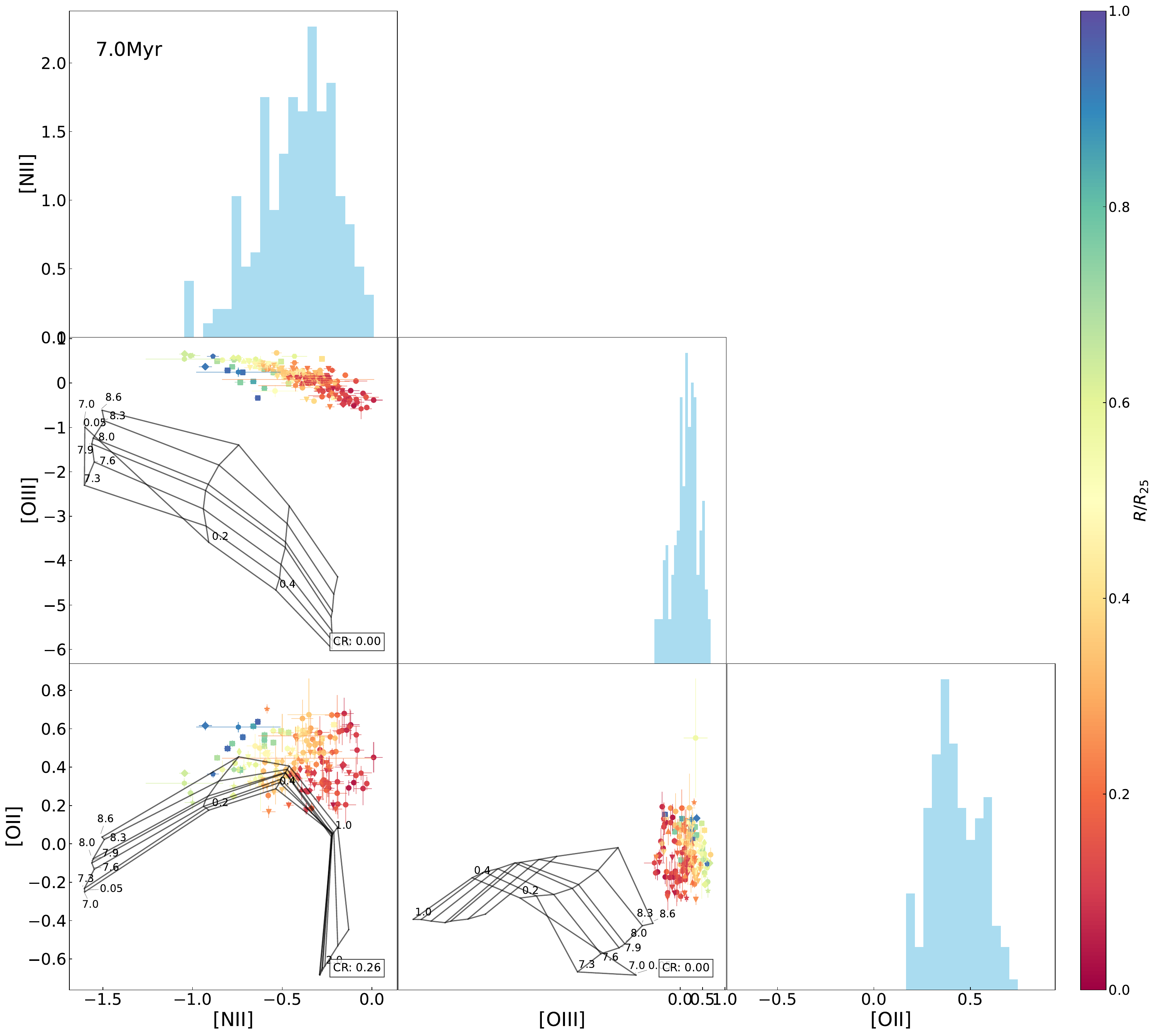}

  \caption{{The} 
 same as Figure~\ref{fig:emission_line_grid_0Myr} but~superimposing the 7.0\,Myr model~grids.}
  \label{fig:emission_line_grid_7.0Myr}
\end{figure}
\unskip

\begin{figure}[H]
  
  \includegraphics[width=0.9\textwidth]{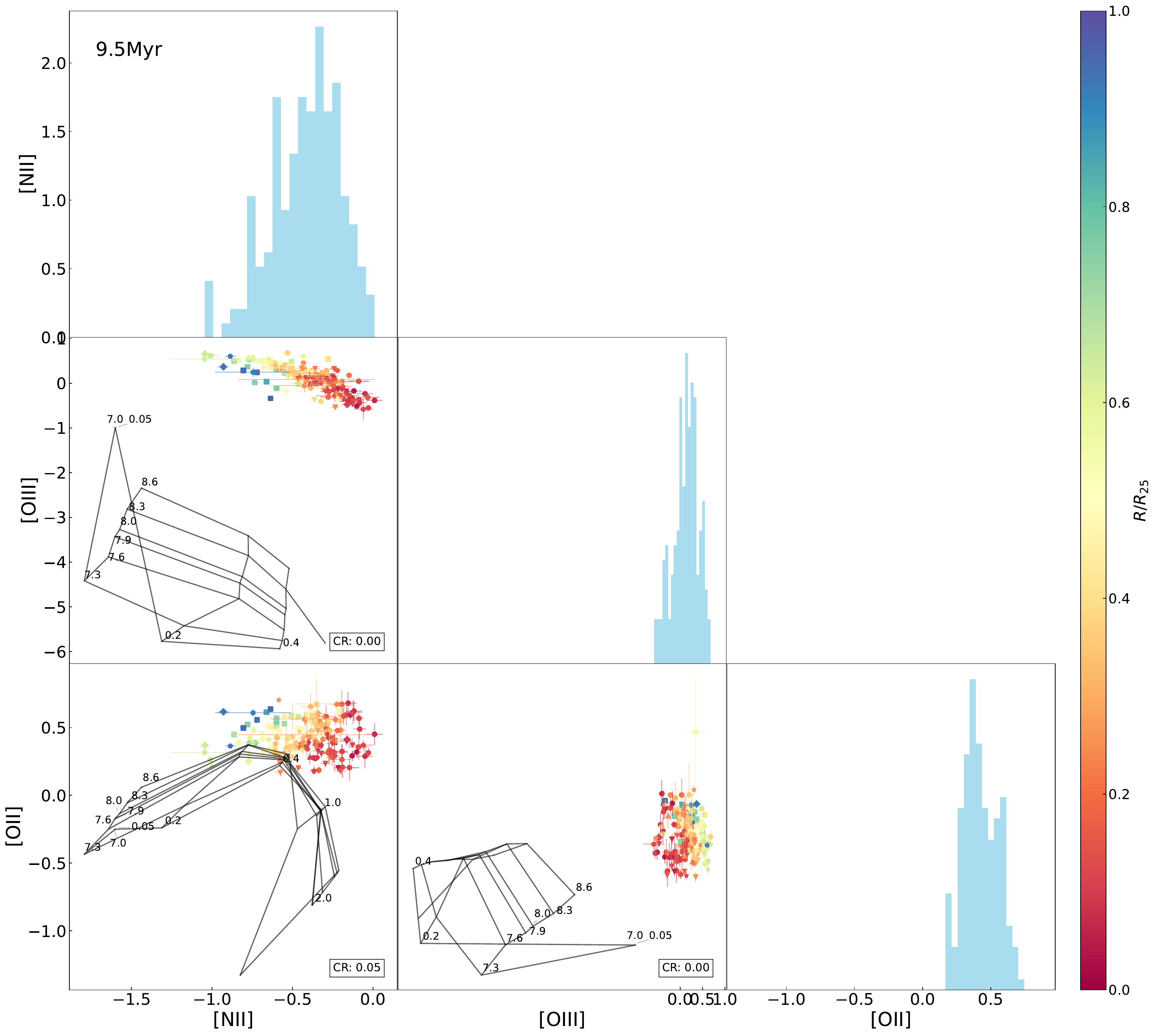}

  \caption{{The} 
 same as Figure~\ref{fig:emission_line_grid_0Myr} but~superimposing the 9.5\,Myr model~grids.}
  \label{fig:emission_line_grid_9.5Myr}
\end{figure}

}

\begin{adjustwidth}{-\extralength}{0cm}


\printendnotes[custom]

\reftitle{References}


\PublishersNote{}
\end{adjustwidth}




\end{document}